\documentclass[twocolumn,aps,prl,superscriptaddress]{revtex4-2}  
\usepackage{graphicx}  
\usepackage{bm}        
\usepackage{amssymb}   
\usepackage{xcolor}
\usepackage{amsmath}
\usepackage{braket}
\definecolor{linkColor}{RGB}{0,80,150}
\usepackage{hyperref}
\usepackage{algpseudocode}
\usepackage{algorithm}

 \usepackage{orcidlink}  

\usepackage{varwidth}

\usepackage[normalem]{ulem}
\usepackage{soul}

\hypersetup{
    colorlinks=true,
    allcolors=linkColor,
    pdfborder={0 0 0},
    pdfencoding = auto
}

\usepackage{xr}
\makeatletter
 
\begin{document}

\title{ Correspondence between quasiparticle dissipation and quantum information decay in open quantum systems}

\author{Zihang Wang\,\orcidlink{0000-0003-4482-117X}}

\email{zihangwang@ucsb.edu}

\affiliation{Department of Physics, University of California Santa Barbara, Santa Barbara, California 93106, USA}
 
 \author{Dirk Bouwmeester\,\orcidlink{0000-0002-2118-6532}}

\email{bouwmeester@ucsb.edu}

\affiliation{Department of Physics, University of California Santa Barbara, Santa Barbara, California 93106, USA}
 \affiliation{Huygens-Kamerlingh Onnes Laboratory, Leiden University, P.O. Box 9504, 2300 RA Leiden, Netherlands}
 
\begin{abstract}
Diagrammatic techniques simplify a weakly interacting many-body problem into an effective few-quasiparticle problem within a system of interest (SOI). If scattering events, mediated by a bath, between those quasiparticles can be approximated as density-density interactions, the bath behaves like an effective external potential. On the other hand, exchange interactions could entangle those quasiparticles and the bath, leading to an open quantum system that induces quantum decoherence and spectral broadening. We investigate the renormalized interaction between the SOI and the bath, employing a projection operator technique similar to the one used in the Nakajima—Zwanzig method \cite{10.1063/1.1731409,breuer2002theory}. We find that the frequency variation of this renormalized interaction is analogous to the quasiparticle residue, and provides a measure of the SOI-bath separability that serves as the lower bound of the SOI-bath entanglement entropy. In the weak coupling regime and continuum limit, we demonstrate that {the degree of SOI-bath separability corresponds to the quasiparticle spectral weight in the single-impurity Anderson model (SIAM), and find that the loss of quantum information to the continuum of the bath can be understood as a decay process where an initial single impurity state escapes to a thermal bath.} This work opens a new direction for connecting energy dissipation in quasiparticles propagation to the loss of quantum information in open quantum systems.

\end{abstract}

\maketitle

\definecolor{lightblue}{RGB}{200,220,255}
\setlength{\fboxsep}{0.03\linewidth}
\noindent\fcolorbox{white}{white}{\parbox{0.94\linewidth}{%
}}
\vspace{0.5em}

\begin{figure*}[t!]
    \centering
    \includegraphics{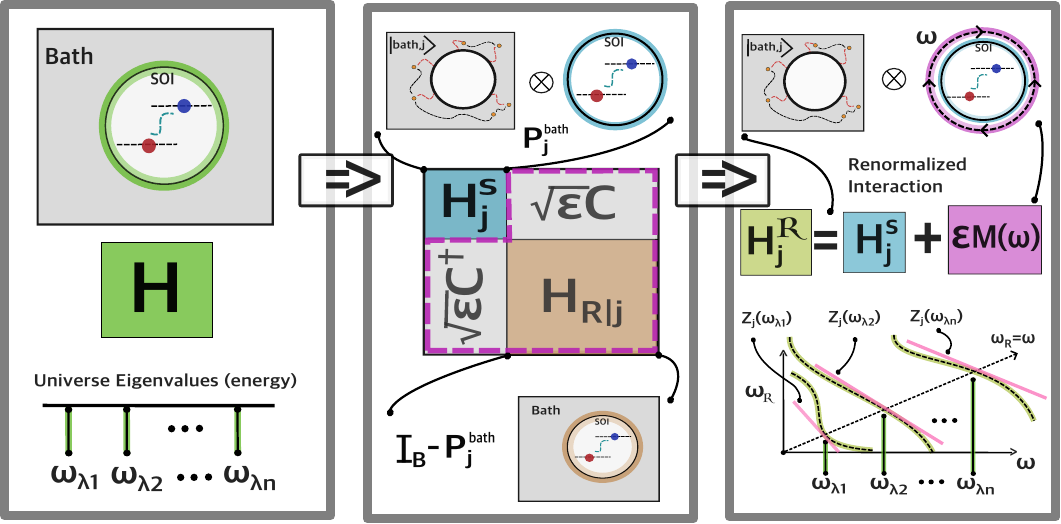}
    \caption{ The panel on the left shows the ``Universe" of the combined System of Interest (SOI) interacting with a bath. The universe has exact energy, $\omega_{\lambda_1},\omega_{\lambda_2},\dots,\omega_{\lambda_n}$ and corresponding eigenstates. In the middle panel a specific (but arbitrary) bath state $\ket{\rm bath, j}$ is selected and the dynamics of the SOI at this given bath state is described by the static Hamiltonian $\mathbf{H}^{\mathcal{S}}_{j}$ where the label S stands for static because the single bath state $\ket{\rm bath, j}$ and SOI form a closed quantum system, as described in Eq.\ref{Hamiltonian projection}. The full dynamics of the universe is re-established by including $\mathbf{H}_{R|j}$, describing the rest space consisting of the SOI interaction with the bath minus the selected bath state $\ket{\rm bath, j}$, together with the coupling term $C$ between the two spaces. By including a perturbation parameter $\epsilon$ with a value between 0 and 1 we can gradually turn on the coupling. The result of the coupling to the rest space can be incorporated as a renormalized interaction $\epsilon\mathbf{M}(\omega)$ as illustrated in the right panel. The renormalized Hamiltonian $\mathbf{H^{\boldsymbol{{\mathcal{R}}}}_j}$ recovers the universe eigenvalues $\omega_{\lambda_1},\omega_{\lambda_2},\dots,\omega_{\lambda_n}$ by the nonlinear fixed point equations $\omega_\mathcal{R}=\omega$ for the subsystem of SOI and bath state $\ket{\rm bath, j}$, as discussed in Eq.~\ref{fixed point equation}. A main result of this article is the observation that the variation around each fixed point, indicated by the pink lines in the graph on the bottom right, captures the universe's eigenstate separability, $\mathbf{Z_j}(\omega_\lambda)$. }
    
    \label{fig demo}
\end{figure*}

\section{Introduction}
The propagation of a few bare particles in a many-body ensemble can be simplified diagrammatically through the self-energy, reducing the many-body complexity to the propagation of a few weakly interacting quasiparticles \cite{anderson2019basic}. Although this is exact, in practice, the self-energy involves approximations that trace out the bath degrees of freedom (DOF) while retaining essential DOF from the system of interest (SOI). The concept of a finite quasiparticle lifetime is related to the inherent damping or decoherence in the presence of the bath, closely linked to the dissipative dynamics of open quantum systems. Various attempts have been made to bridge the two: For example, in field theory, the dynamics of a dissipative two-state system, governed by the spin-boson Hamiltonian, can be described through a path integral approach \cite{RevModPhys.59.1}; in a more practical approach, dynamical mean-field theory (DMFT) \cite{RevModPhys.86.779,RevModPhys.68.13,RevModPhys.78.865}, a state-of-the-art {ab initio} technique for understanding strongly correlated materials, involves mapping the evolution of a one- or two-particle propagator to an equivalent open quantum system known as the impurity Anderson model (IAM) \cite{martin_reining_ceperley_2016};  with an identical bath correlation function, the pseudomode method introduces a few ``unphysical'' harmonic (pseudo) modes with dissipative coupling to the SOI, governed by an equivalent master equation \cite{PhysRevA.55.2290,Lambert2019}; in the strong SOI-bath coupling regime, the dynamics of a dissipative SOI can be mapped to an effective Hamiltonian with dressed interactions \cite{PRXQuantum.4.020307}.

One common feature of open quantum systems is the inseparability of the ``universe'' (SOI plus bath), as for example, quantified by the Peres–Horodecki criterion \cite{PhysRevLett.77.1413}. The violation of this criterion implies non-local correlations that cannot be described by local hidden-variable theories \cite{RevModPhys.65.803,PhysicsPhysiqueFizika.1.195}. As a result, the reconstruction of the universe through local operations and classical communications is not possible. The non-local correlations imply non-classical interactions, resulting in a mixed reduced density matrix of the SOI \cite{RevModPhys.88.021002}. A precise measure of the non-local correlations is given by the minimum of all basis-dependent quantum discords, which exactly recovers the entanglement entropy in case of a pure state \cite{PhysRevA.79.042325, PhysRevLett.105.020503,PhysRevLett.88.017901,PhysRevLett.105.190502}. Temporal correlations, as probed by sequential measurements, can also be explored to quantify the quantum nature of a system, even in the absence of entanglement  \cite{10.3389/fphy.2024.1325239}.

The evolution dynamics of the density matrix, $\boldsymbol{\rho_\lambda}=\ket{\boldsymbol{\lambda}}\bra{\boldsymbol{\lambda}}$, constructed from an energy eigenstate of the ``universe'' is principle trivial but inaccessible because of the huge Hilbert space dimension. However, quantifying the dynamics of the reduced density matrix, $\boldsymbol{\rho_{\rm SOI}}=\mathrm{Tr}_{\rm bath}\left(\ket{\boldsymbol{\lambda}}\bra{\boldsymbol{\lambda}}\right)$, is challenging and usually involves heavy approximations such as the Born-Markov approximation in the corresponding master equations, e.g. the decay term in the Lindblad form, $[\boldsymbol{\rho_{\rm SOI}},H_I(t)] \to \mathcal{L}(t)\boldsymbol{\rho_{\rm SOI}}$, where $H_I(t)$ is the interaction Hamiltonian and $\mathcal{L}(t)$ is the Liouville super-operator \cite{10.1063/1.5115323,RevModPhys.94.045006}, or with hierarchical equations of motions \cite{doi:10.1143/JPSJ.74.3131, doi:10.1143/JPSJ.75.082001}. 

 The reduction in bath DOF often leads to a non-Markovian evolution of the reduced density operator of the SOI \cite{Nazir2018,PhysRevA.104.052617,RevModPhys.94.045006,breuer2002theory}, and can be captured by techniques like the Nakajima—Zwanzig method or time-convolutionless technique. Both techniques utilize fixed bath states, such as the stationary Gibbs state of the bath, to generate a closed master equation for the reduced density matrix of the SOI \cite{breuer2002theory, 10.1063/1.1731409}. {Recently, it has been experimentally demonstrated that non-Markovian coupling could lead to memory effects where the quantum information could be further retrieved from the bath \cite{PhysRevLett.132.200401}. This adds another incentive to go beyond the Markov limit where the time (frequency) dependence bath degrees of freedom becomes essential. }

Despite advancements in approximating the excitation spectrum and dynamics of open quantum systems, the link between the energy eigenstate of the ``universe" and the corresponding effective eigenstate of the open quantum system remains less clear: How well can we approximate the bath-SOI interaction at a particular excitation energy as classical interaction, described by a renormalized density-density interaction? In other words, at the given energy, can we model the interaction as a many-body ensemble residing in an effective external potential?

In this work, we address this question by examining the renormalization effects between a SOI and a bath. We first demonstrate that, for a given bath state, the frequency variation of the renormalized interaction with the SOI captures a lower bound of the SOI-bath entanglement entropy, a measure of the universe eigenstate separability. By sampling various bath states, this lower bound provides a quantitative measure that reflects the validity of approximating the bath-SOI interaction as density-density interaction at a given energy. Subsequently, we prove that in the weak coupling limit, this lower entropy bound exhibits a frequency dependence that can be derived from the spectrum and weights of a general single-impurity Anderson model (SIAM). This direct correspondence establishes a theoretical foundation for connecting the quantum information decay in open quantum systems with effective quasiparticle energy dissipation.

\section{ Formalism of Renormalization}
Let's consider a non-degenerated many-body Hamiltonian $\mathbf{H}$ that has $\Lambda$ (integer) {unique} eigenvalues $\omega_\lambda$ and corresponding eigenstates $\ket{\boldsymbol{\lambda}} $, $ \mathbf{H} \ket{\boldsymbol{\lambda}}=\omega_{\lambda}\ket{\boldsymbol{\lambda}}$, as illustrated in the left panel of Fig.~\ref{fig demo}, (we set $\hbar=1$). The {physical} basis ${\ket{i}}$ that spans the corresponding universe's Hilbert space $\mathcal{H}=\mathcal{H}_{\rm SOI} \otimes \mathcal{H}_{\rm bath}$, is partitioned into two distinguishable spaces, often labeled by their spins, polarizations, occupation number, angular momentum, etc. We refer to them as the Hilbert spaces of the SOI ($\mathcal{H}_{\rm SOI}$, dimension $\mathcal{D}_{\rm SOI} $) and of the bath ( $\mathcal{H}_{\rm bath}$, dimension $\mathcal{D}_{\rm bath} $), with, $\mathcal{D}_{\rm SOI} \leq \mathcal{D}_{\rm bath}$. Each universe's eigenstate therefore can be expanded in the system and bath bases, $\{\ket{\rm SOI, i}\}$ and $\{\ket{\rm bath, j}\}$,  
\begin{equation}
\ket{\boldsymbol{\lambda}}=\sum_{i,j}^{\mathcal{D}_{\rm SOI},\mathcal{D}_{\rm bath}} w^{ij}_\lambda
   \ket{\rm SOI, i} \otimes  \ket{\rm bath, j} ,
  \label{physical state1}
\end{equation}
where $ w^{ij}_\lambda$ are complex-valued coefficients. As illustrated in the middle panel of Fig.~\ref{fig demo}, let's consider a {static} Hamiltonian $\mathbf{H}^{\mathcal{S}}_j$ that incorporates static interactions between a particular bath state $\ket{\rm bath,j}$ and the SOI, 
\begin{equation}
\begin{aligned}
    &\mathbf{H}^{\mathcal{S}}_{j} =\mathbf{P}^{\rm bath}_{j} \mathbf{H}  \mathbf{P}^{\rm bath}_{j},
\end{aligned}
    \label{Hamiltonian projection}
\end{equation}
where the projection operator,
\begin{equation}
    \mathbf{P}^{\rm bath}_{j} = \ket{\rm bath, j} \bra{\rm bath, j}, \hspace{0.2cm} \sum_j  \mathbf{P}^{\rm bath}_{j}=\mathbf{I_B},
    \label{projector}
\end{equation} selects interactions involving a specific bath state $\ket{\rm bath,j}$ (a pointer state). The {rest space} Hamiltonian $\mathbf{H}_{R|j}$,
\begin{equation}
    \mathbf{H}_{R|j}=\left(\mathbf{I_B}-\mathbf{P}^{\rm bath}_{j}\right)  \mathbf{H}  \left(\mathbf{I_B}-\mathbf{P}^{\rm bath}_{j}\right),
\end{equation}
contains all remaining configuration interactions. We denote the Hilbert space after the bath state projection as the {projected} Hilbert space, $\mathcal{H}_{\rm P_j}=\mathcal{H}_{\rm SOI} \otimes \{\ket{\rm bath,j}\}$. {The coupling between states in the projected Hilbert space and the rest space, is denoted by $\mathbf{C}$:}
\begin{equation}
    \mathbf{C}=\mathbf{P}^{\rm bath}_{j} \mathbf{H} \left(\mathbf{I_B}-\mathbf{P}^{\rm bath}_{j}\right), \hspace{0.2cm}  \mathbf{C^\dag}= \left(\mathbf{I_B}-\mathbf{P}^{\rm bath}_{j}\right) \mathbf{H} \mathbf{P}^{\rm bath}_{j}. 
\end{equation}

In the static Hamiltonian $\mathbf{H}^{\mathcal{S}}_{j}$, the bath state $\ket{\rm bath, j}$ serves as a fixed ``background potential'', interacting with the SOI {statically} solely through the density interaction, while interactions within the SOI remains quantum mechanical. The corresponding energy eigenvalue equation is, $\mathbf{H}_j^{\mathcal{S}} \ket{\boldsymbol{\mathcal{S}_j}} = \omega_{\mathcal{S}} \ket{\boldsymbol{\mathcal{S}_j}}$, where the eigenstate $\ket{\boldsymbol{\mathcal{S}_j}}$ lives on the {projected} Hilbert space, therefore it is a product state (or a simply separable state) of $\ket{\rm bath,j}$ and one state from the SOI. The choice of $\ket{\rm bath,j}$ is arbitrary yet essential for quantifying the validity of approximating the interaction as a classical interaction near a given excitation energy. {Although the bath is commonly considered as a continuum such as a thermal ensemble, the bath states can be discrete.} 

{In a matrix form, the eigenvalue equation of the full Hamiltonian can be expressed as two coupled equations that contains the static Hamiltonian and rest space Hamiltonian (see Fig.~\ref{fig demo} middle panel),}
\begin{equation}
\begin{aligned}
  &\omega_{\lambda}\ket{ \boldsymbol{\lambda}} =\mathbf{H}   \ket{ \boldsymbol{\lambda}} = \begin{bmatrix}
       \mathbf{H}^{\mathcal{S}}_{j} & \sqrt{\epsilon}\mathbf{C}\\
     \sqrt{\epsilon} \mathbf{C}^\dag & \mathbf{H}_{R|j}
  \end{bmatrix} \left(a^j_\lambda\ket{\boldsymbol{{\mathcal{R}}_j}}\oplus b^j_\lambda\ket{\boldsymbol{{\mathcal{R}}_{R|j}}}\right),\\
\end{aligned}
\label{general H}
\end{equation}
where $\epsilon$ is a dimensionless perturbation parameter ranging from 0 to 1. When $\epsilon=0$, the SOI is detached from the bath degrees of freedom. The coupling block $\mathbf{C}$ is a frequency independent rectangular matrix that captures the interaction between the two subsystems. We rewrite the universe's eigenstates as {direct sums } of two projectors, $a^j_\lambda \ket{\boldsymbol{\mathcal{R}_j}} = \mathbf{P}^{\rm bath}_{j} \ket{\boldsymbol{\lambda}} $ and $b^j_\lambda \ket{\boldsymbol{{\mathcal{R}}_{R|j}}} = (\mathbf{I_B}-\mathbf{P}^{\rm bath}_{j}) \ket{\boldsymbol{\lambda}} $ with coefficients $a^j_\lambda$ and $b^j_\lambda$ that normalize to 1. {The above equation can be further expanded as,
\begin{equation}
\begin{aligned}
     \begin{bmatrix}
           \mathbf{H}^{\mathcal{S}}_{j} \\
          \sqrt{\epsilon}\mathbf{C}  \\
         \end{bmatrix} & \cdot a^j_\lambda \ket{\boldsymbol{\mathcal{R}_j}}  = [\mathbf{P}^{\rm bath}_{j} \mathbf{H}  \mathbf{P}^{\rm bath}_{j} \\
         &\oplus  \sqrt{\epsilon} \mathbf{P}^{\rm bath}_{j} \mathbf{H} \left(\mathbf{I_B}-\mathbf{P}^{\rm bath}_{j}\right) ] \mathbf{P}^{\rm bath}_{j} \ket{\boldsymbol{\lambda}}  \\
         &=a^j_\lambda \mathbf{H}^{\mathcal{S}}_{j} \ket{\boldsymbol{{\mathcal{R}}_j}}.
\end{aligned}
\end{equation} 
}
and in components,
\begin{equation}
  \left\{ \begin{aligned}
&\omega_{\lambda}a^j_\lambda\ket{\boldsymbol{{\mathcal{R}}_j}}=  \mathbf{H}^{\mathcal{S}}_{j} a^j_\lambda\ket{\boldsymbol{{\mathcal{R}}_j}} \oplus \sqrt{\epsilon}\mathbf{C} b^j_\lambda\ket{\boldsymbol{{\mathcal{R}}_{R|j}}}\\
&\omega_{\lambda}b^j_\lambda\ket{\boldsymbol{{\mathcal{R}}_{R|j}}}=  \sqrt{\epsilon}\mathbf{C^\dag} 
 a^j_\lambda\ket{\boldsymbol{{\mathcal{R}}_j}} \oplus \mathbf{H}_{R|j} b^j_\lambda\ket{\boldsymbol{{\mathcal{R}}_{R|j}}}
\end{aligned} \right. .
\end{equation}

Since we only consider the dynamics in the SOI with a given bath state $\ket{\rm bath, j}$, we rewrite the above coupled eigenvalue equations into a non-linear eigenvalue equation by replacing the $\ket{\boldsymbol{{\mathcal{R}}_{R|j}}}$ with $\ket{\boldsymbol{\mathcal{R}_j}}$ via the relation,  
 \begin{equation}
\begin{aligned}
    &b^j_\lambda\ket{\boldsymbol{{\mathcal{R}}_{R|j}}}= \sqrt{\epsilon} 
 \frac{1}{\omega_{\lambda}-\mathbf{H}_{R|j}} \mathbf{C}^\dag \ket{\boldsymbol{{\mathcal{R}}_j}}a^j_\lambda,\\&(b^j_\lambda)^*\bra{\boldsymbol{{\mathcal{R}}_{R|j}}}=(a^j_\lambda)^* \bra{\boldsymbol{{\mathcal{R}}_j}} \mathbf{C} \frac{1}{\omega_{\lambda}-\mathbf{H}_{R|j}} \sqrt{\epsilon}  ,
\end{aligned}
\label{first equations}
\end{equation}
 and using the fact that the rest space Hamiltonian is Hermitian, $\mathbf{H}^\dag_{R|j}=\mathbf{H}_{R|j}$. The substitution leads to a renormalized (frequency-dependent) Hamiltonian that lives on the same {projected} Hilbert space as the static Hamiltonian,
\begin{equation}
    \mathbf{H}^{\mathcal{R}}_{j} (\omega) = \mathbf{H}^{\mathcal{S}}_{j}+ \epsilon \boldsymbol{ M}(\omega), \hspace{0.3cm} \boldsymbol{ M}(\omega)=\mathbf{C} \frac{1}{\omega-\mathbf{H}_{R|j}} \mathbf{C}^\dag,
\label{self energy and rn block def}
\end{equation} where $\boldsymbol{M}(\omega)$ is precisely the {Schur complement} of the rest space Hamiltonian \cite{martin_reining_ceperley_2016}. It captures the renormalization effects induced {by the remaining degrees of freedom of the bath}. {Loosely speaking, in the language of scattering theory, the frequency dependence in the renormalized Hamiltonian originates from scattering events that are mediated by bath state interactions \cite{Sakurai_Napolitano_2017,martin_reining_ceperley_2016}.} The eigenvalues and eigenstates of the renormalized Hamiltonian are calculated via the non-linear eigenvalue equation, 
\begin{equation}
    \mathbf{H}^{{\mathcal{R}}}_{j} (\omega) \ket{\boldsymbol{{\mathcal{R}}_j}}=\omega_{{\mathcal{R}}}(\omega)\ket{\boldsymbol{{\mathcal{R}}_j}},
\label{eq:non_lin_eival}
\end{equation}
which has $\Lambda$ fixed point solutions that recover {all} many-body eigenvalues, i.e.,
\begin{equation}
    \omega_\lambda=\omega_{\mathcal{R}}(\omega_\lambda),
    \label{fixed point equation}
\end{equation}
and $\Lambda > \mathcal{D}_{\rm SOI}$. The eigenvalues and eigenstates of the renormalized Hamiltonian are frequency-dependent and only hold {proper} meaning near the fixed point solutions.

Let's summarize the key similarities and differences between a static Hamiltonian $\mathbf{H}^S_j  $ and a renormalized Hamiltonian $\mathbf{H}^{\mathcal{R}}_j  $. Firstly, both Hamiltonians operate in the {same} Hilbert space; thus, by definition, each of their eigenstates must be a product of the bath state $\ket{\rm bath, j}$ and a state from the SOI. Secondly, the renormalized Hamiltonian recovers $\Lambda$ exact universe's eigenvalues via fixed-point equations, while the static Hamiltonian only possesses $\mathcal{D}_{\rm SOI}$ static eigenvalues. For notation convenience, we define a transform {operator} acting on eigenstates of the static Hamiltonian that produces eigenstates of the renormalized Hamiltonian at a given energy,
\begin{equation}
\ket{\boldsymbol{\mathcal{R}_j}} = \mathcal{U}_j(\epsilon,\omega_{\mathcal{S}\lambda})\ket{\boldsymbol{\mathcal{S}_j}}, \hspace{0.4cm} \omega_{\mathcal{S}\lambda}=\omega_{\lambda}-\omega_{\mathcal{S}}, 
\label{transform operator}
\end{equation}
where $\omega_{\mathcal{S}\lambda}$ is the energy detuning, and unitarity gives $\mathcal{U}^\dag_j(\epsilon,\omega_{\mathcal{S}\lambda})\mathcal{U}_j(\epsilon,\omega_{\mathcal{S}\lambda})=\mathbf{I_{\rm SOI}} \otimes     \mathbf{P}^{\rm bath}_{j} $. The identity operator of the universe can then be written as, $\mathbf{I}=\mathbf{I_{\rm SOI}} \otimes \mathbf{I_B} $. Since $\ket{\boldsymbol{\mathcal{S}_j}}$ are non-degenerate eigenstates of the static Hamiltonian, unitary operators can be uniquely specified by the energy detuning $\omega_{\mathcal{S}\lambda}$.

\section{renormalized eigenstates and entanglement entropy }
From a practical perspective, if our primary goal is to recover the universe eigenvalues, solving the renormalized Hamiltonian is sufficient \cite{Gao2016, Ugeda2014,DESLIPPE20121269}. However, it is important to note that the eigenstates of the renormalized Hamiltonian contain entanglement information, which has been overlooked so far. As an example, let's consider a scenario of two excitons (two sets of electron-hole pairs) in a box that interact quantum mechanically. We address the question: How well can we approximate the system as a product state of two excitons with renormalized interactions at a given energy? Following the discussion above, we can choose an arbitrary electron-hole configuration as the bath state (i.e., a fixed exciton), while keeping the remaining electron-hole degrees of freedom as the system of interest (SOI). Subsequently, we can define the renormalized Hamiltonian and recover the energy spectrum exactly. However, describing the above system as a product state, which is usually done in the Born-Markov approximation, may not always be appropriate. Such a product state only capture {partial} information of the universe's eigenstates, and the difference can be related to the lower bound of the SOI-bath entanglement entropy, {The corresponding universe's von Neumann entropy is defined as the trace over all bath degrees of freedom, 
\begin{equation}
    \mathcal{E}(\omega_\lambda)=-\mathrm{Tr}\left(\boldsymbol{\rho_{\rm SOI}} \mathrm{Log}(\boldsymbol{\rho_{\rm SOI}})\right), \hspace{0.2cm} \boldsymbol{\rho_{\rm SOI}}=\mathrm{Tr_{bath}}\left(\boldsymbol{\rho_\lambda}\right).
\end{equation}
 }
a quantity that reflects the degree of entanglement between the SOI and the bath.

Let's first examine the overlap between a universe eigenstate and an eigenstate from the renormalized Hamiltonian near a fixed point, denoted as the degree of separability, $\boldsymbol{{Z}_j}(\omega_{\lambda}) \equiv \braket{\boldsymbol{\mathcal{R}_j} | \boldsymbol{\rho_\lambda} |\boldsymbol{\mathcal{R}_j} }=|a^j_\lambda|^2$, ranging from $0$ to 1. It can be rewritten via the eigenstate of the static Hamiltonian, 
\begin{equation}
\begin{aligned}
       \boldsymbol{{Z}_j}(\omega_{\lambda}) &= \bra{\boldsymbol{\mathcal{S}_j}}\mathcal{U}^\dag_j(\epsilon,\omega_{\mathcal{S}\lambda}) \boldsymbol{\rho_\lambda} \mathcal{U}_j(\epsilon,\omega_{\mathcal{S}\lambda})\ket{\boldsymbol{\mathcal{S}_j}}.\\
\end{aligned}
\end{equation}
{If $\boldsymbol{{Z}_j}(\omega_{\lambda}) =1$, the universe's eigenstate $\ket{\boldsymbol{\lambda}}$ is fully separable, i.e. a product state. If $\boldsymbol{{Z}_j}(\omega_{\lambda}) <1$, by the normalization condition, the remain weights must transferred to the remaining degrees of freedom that are orthogonal to the bath state $\ket{\rm bath,j}$. The simplest case for the universe's eigenstate involves the following sum,
\begin{equation}
  \boldsymbol{\rho^{(2)}_\lambda}=\boldsymbol{{Z}_j}(\omega_{\lambda}) \ket{\boldsymbol{\mathcal{R}_j} }\bra{\boldsymbol{\mathcal{R}_j} }+(1-\boldsymbol{{Z}_j}(\omega_{\lambda}) )\ket{\boldsymbol{\mathcal{R}_{j'}} }\bra{\boldsymbol{\mathcal{R}_{j'}} },
\end{equation}
where the density matrix for a universe's eigenstate $\boldsymbol{\rho^{(n)}_\lambda}$ with $\mathbf{n}=2$ indicates that this particular eigenstate is the sum of two product states. In other words, the eigenstate of the rest space Hamiltonian, $\ket{\boldsymbol{{\mathcal{R}}_{R|j}}}$ is a product state that only includes a single bath state $\ket{\rm bath,j'}$ that is orthogonal to the selected bath state $\ket{\rm bath,j}$, i.e. $\braket{\boldsymbol{\mathcal{R}_j} |\boldsymbol{\mathcal{R}_{j'}} }=0$. The entanglement entropy of the above universe's eigenstate $\boldsymbol{\rho^{(2)}_\lambda}$, is given as, 
\begin{equation}
   B(\boldsymbol{{Z}_j})= -\boldsymbol{{Z}_j} \mathrm{Log}[\boldsymbol{{Z}_j} ]-(1-\boldsymbol{{Z}_j})  \mathrm{Log}[1-\boldsymbol{{Z}_j}],
   \label{lower bound equation}
\end{equation}
where we define $B(\boldsymbol{{Z}_j})=-\mathrm{Tr}\left(\boldsymbol{\rho^{(2)}_{\rm SOI}} \mathrm{Log}(\boldsymbol{\rho^{(2)}_{\rm SOI}})\right)$ with $\boldsymbol{\rho^{(2)}_{\rm SOI}}=\mathrm{Tr_{bath}}\left(\boldsymbol{\rho^{(2)}_\lambda}\right)$.}

{As discussed in the supplementary material, the degree of separability establishes the lower bound of the entropy of the universe at particular energy $\omega_\lambda$, i.e. $ \mathcal{E}(\omega_{\lambda}) \geq B(\boldsymbol{{Z}_j})$. Intuitively, this lower bound can be understood as the minimal way to distribute the weights over a given universe's eigenstate $\boldsymbol{\rho^{(n)}_\lambda}$: As we allow more bath degrees of freedom $\mathbf{n} \geq 2$, the entropy increases caused by the increase in the universe's multiplicity \cite{PhysRevA.92.032316}.}

{To gain more physical insights on the degrees of separability, let's revisit the renormalized Hamiltonian $  \mathbf{H}^{\mathcal{R}}_{j} (\omega)$ defined in Eq.~\ref{eq:non_lin_eival}. Since it is frequency dependent, there exists a family of eigenstates $\ket{\boldsymbol{{\mathcal{R}}_j}}$ and eigenvalues $\omega_{{\mathcal{R}}}(\omega)$ that satisfies the non-linear equation, Eq.~\ref{eq:non_lin_eival} apart from the fixed point equation Eq.~\ref{fixed point equation}. The eigenvalues of Eq.~\ref{eq:non_lin_eival}, $\omega_{{\mathcal{R}}}(\omega)$ defines a family of {interaction curves} that {smoothly} connects fix points, i.e. $\omega_{\lambda}=\omega_{{\mathcal{R}}}(\omega_\lambda)$. Although those frequencies in Eq.~\ref{eq:non_lin_eival} that do {not} satisfy the fixed point equation have no direct physical meaning, near each fixed point, the variation, ranged from $(-\infty,0]$, directly recovers the degree of separability,}
\begin{equation}
\begin{aligned}
       &  \left. \frac{\partial \omega_{{\mathcal{R}}}(\omega) }{\partial \omega}\right|_{\omega_{\lambda }} =  \epsilon \bra{\boldsymbol{{\mathcal{R}}_j}}\mathbf{C} \frac{\partial}{\partial \omega} \left. \left(  \frac{1}{\omega- \mathbf{H}_{R|j}}\right)\right|_{\omega_{\lambda }} \mathbf{C}^\dag \ket{\boldsymbol{{\mathcal{R}}_j}}\\
         &=-\epsilon\bra{\boldsymbol{{\mathcal{R}}_j}} \mathbf{C}  \left. \left(  \frac{1}{\omega_{\lambda}-\mathbf{H}_{R|j}}\right)\left(  \frac{1}{\omega_{\lambda}-\mathbf{H}_{R|j}}\right)\right. \mathbf{C}^\dag \ket{\boldsymbol{{\mathcal{R}}_j}}
       \\
       &= - \frac{|b^j_\lambda|^2}{|a^j_\lambda|^2}=1-\frac{1}{\boldsymbol{{Z}_j}(\omega_{\lambda})}. \\
\end{aligned}
\label{fixed point variation}
\end{equation}
In other words, the degree of separability $\boldsymbol{{Z}_j}(\omega)$ has an alternative representation near each fixed point, as given in Eq.~\ref{fixed point variation}, 
\begin{equation}
\boldsymbol{{Z}_j} (  \omega_{\lambda}) = \left(1-\left. \frac{\partial \omega_{{\mathcal{R}}}(\omega) }{\partial \omega}\right|_{\omega_{\lambda }} \right)^{-1}. 
    \label{grad curve}
\end{equation}
It is analogous to the quasiparticle residue, derived from the variation of the self-energy \cite{PhysRevB.91.121107,martin_reining_ceperley_2016}. {The above relation implies that the lower bound of the entanglement entropy $B[\omega_\lambda]$ can be approximated by examining the energy variation in the vicinity of a universe eigenvalue. It is evident that when the variation is $-1$, the lower bound gives $B=\mathrm{Log}(2)$.}

{We define the frequency dependent weight factor centered at the eigenvalue $\omega_{\mathcal{S}}$ of the static Hamiltonian},
\begin{equation}
    \boldsymbol{\mathcal{W}_j} ( \omega-\omega_{\mathcal{S}})=\boldsymbol{{Z}_j} (   \omega )\mathbf{z}_{j}(  \omega-\omega_{\mathcal{S}}),
\end{equation}
where we define the degree of similarity, $\mathbf{z}_{j}(\omega_{\mathcal{S}\lambda})\equiv\braket{\boldsymbol{\mathcal{R}_j} | \boldsymbol{\rho_\mathcal{S}} |\boldsymbol{\mathcal{R}_j} }$, or equivalently, 
\begin{equation}
\mathbf{z}_{j}(\omega_{\mathcal{S}\lambda}) =\bra{\boldsymbol{\mathcal{S}_j}} \mathcal{U}^\dag_j(\epsilon,\omega_{\mathcal{S}\lambda}) \boldsymbol{\rho_{\mathcal{S}}}\mathcal{U}_j(\epsilon,\omega_{\mathcal{S}\lambda}) \ket{\boldsymbol{\mathcal{S}_j}},
\end{equation}
{and $\boldsymbol{\rho_{\mathcal{S}}}=\ket{\boldsymbol{\mathcal{S}_j}} \bra{\boldsymbol{\mathcal{S}_j}} $ is the static density matrix with energy $\omega_{\mathcal{S}}$.}
When evaluating the weight factor at each fixed point, we recover $|\braket{\boldsymbol{\mathcal{S}_j} | \boldsymbol{ \lambda}}|^2 = \boldsymbol{\mathcal{W}_j} ( \omega_\lambda-\omega_\mathcal{S})$. It can be interpreted as the {single-particle} spectral weight, representing the probability density of observing a bare particle with energy $\omega_{\lambda}$ in a many-body problem. It satisfies $\lim_{\epsilon\to 0} \boldsymbol{\mathcal{W}_j}= 1$. {The sum rule of the weight factor enforces the conservation of probability}, 
\begin{equation}
 1=  \braket{\boldsymbol{{\mathcal{S}}_j}|\boldsymbol{{\mathcal{S}}_j}} = \sum_{  \mathbf{\lambda}} \braket{\boldsymbol{{\mathcal{S}}_j}|\boldsymbol{\lambda}}\braket{\boldsymbol{\lambda}|\boldsymbol{{\mathcal{S}}_j}}=\sum_{ \mathbf{\lambda}} \boldsymbol{\mathcal{W}_j} ( \omega_\lambda-\omega_{\mathcal{S}}),
 \label{sum w}
\end{equation}
{where we insert the resolution of identity of the universe. }

Evaluating the weight factor becomes increasingly challenging as the bath dimension grows. The complexity arises from $ \boldsymbol{{Z}_j}(\omega)$, which appears to require detailed information about the exponentially growing rest space Hamiltonian $\mathbf{H}_{R|j}$. {The degree of separability near a fixed point $\omega_{\lambda}$ can be further expressed into a more instructive form,}
\begin{equation}
 \begin{aligned}
 &\frac{1}{\boldsymbol{{Z}}_j (\omega_\lambda)}-1=\epsilon \bra{\boldsymbol{{\mathcal{R}}_j}}\mathbf{C}\frac{1}{\omega_{\lambda}-\mathbf{H}_{R|j}} \frac{1}{\omega_{\lambda}-\mathbf{H}_{R|j}} \mathbf{C}^\dag \ket{\boldsymbol{{\mathcal{R}}_j}}\\
  &= \frac{1}{\epsilon} \bra{\boldsymbol{{\mathcal{R}}_j}}\mathbf{C}\frac{1}{\omega_{\lambda}-\mathbf{H}_{R|j}} \mathbf{C}^\dag\epsilon  \frac{1}{\boldsymbol{\mathcal{K}}} \epsilon \mathbf{C} \frac{1}{\omega_{\lambda}-\mathbf{H}_{R|j}} \mathbf{C}^\dag \ket{\boldsymbol{{\mathcal{R}}_j}}\\
  &=\frac{1}{\epsilon}\bra{\boldsymbol{{\mathcal{R}}_j}}  (\omega_{\lambda}-\mathbf{H}^{\boldsymbol{\mathcal{S}}}_j ) (\frac{1}{ \boldsymbol{\mathcal{K}}})(\omega_{\lambda}-\mathbf{H}^{\boldsymbol{\mathcal{S}}}_j )\ket{\boldsymbol{{\mathcal{R}}_j}},\\
 \end{aligned}
 \label{overlap general}
\end{equation}
where the coupling kernel $\boldsymbol{\mathcal{K}}$, is an invertible and positively defined matrix with the condition,
\begin{equation}
    \mathbf{C^\dag ({1}/{\boldsymbol{\mathcal{K}}})C=\mathbf{I_{\rm SOI}} \otimes     \mathbf{P}^{\rm bath}_{j} }.
    \label{kernel}
\end{equation} {Since the static Hamiltonian is Hermitian, the eigenstates form an orthonormal basis for the projected Hilbert space,
i.e.,
\begin{equation}
   \mathbf{I_{\rm SOI}} \otimes     \mathbf{P}^{\rm bath}_{j} =\sum_{\mathcal{S}} \ket{\boldsymbol{{\mathcal{S}}_j}}\bra{\boldsymbol{{\mathcal{S}}_j}}.
   \label{idenity}
\end{equation}
Let's simplify Eq.~\ref{overlap general} further by inserting Eq.~\ref{idenity},}
\begin{equation}
\begin{aligned}
 \frac{1}{\boldsymbol{{Z}}_j (\omega_\lambda)}-1&= \sum_{{\mathcal{S}}, {\mathcal{S}}'} \braket{\boldsymbol{{\mathcal{R}}_j}|\boldsymbol{{\mathcal{S}}'_j}}\braket{\boldsymbol{{\mathcal{S}}_j}|\boldsymbol{{\mathcal{R}}_j}} \frac{(\omega_{\lambda}-\omega_{\mathcal{S}} )(\omega_{\lambda}-\omega_{{\mathcal{S}}'} )}{\epsilon\boldsymbol{\mathcal{K}}_{{\mathcal{S}} {\mathcal{S}}'}},\\
\end{aligned}
 \label{equality}
\end{equation}
{and we define the matrix element $  (\boldsymbol{\mathcal{K}}_{{\mathcal{S}} {\mathcal{S}}'})^{-1}=  \bra{\boldsymbol{{\mathcal{S}}_j}}(\frac{1}{ \boldsymbol{\mathcal{K}}})\ket{\boldsymbol{{\mathcal{S'}}_j}}$ for the coupling kernel, which has a direct and intuitive physical meaning: it reflects the transition amplitude between two eigenstates of the static Hamiltonian by considering {all} transition channels induced by the presence of the bath degrees of freedom. Eq.~\ref{equality} is exact, and we will numerically investigate an example in the following section. In the weak interaction limit, $\epsilon \approx 0$, we demonstrated that in the continuum limit, the degree of separability around an eigenvalue of the static Hamiltonian can be understood in terms of the probability of quasi-particle decay. }

\begin{figure*}[t!]
    \centering
    \includegraphics{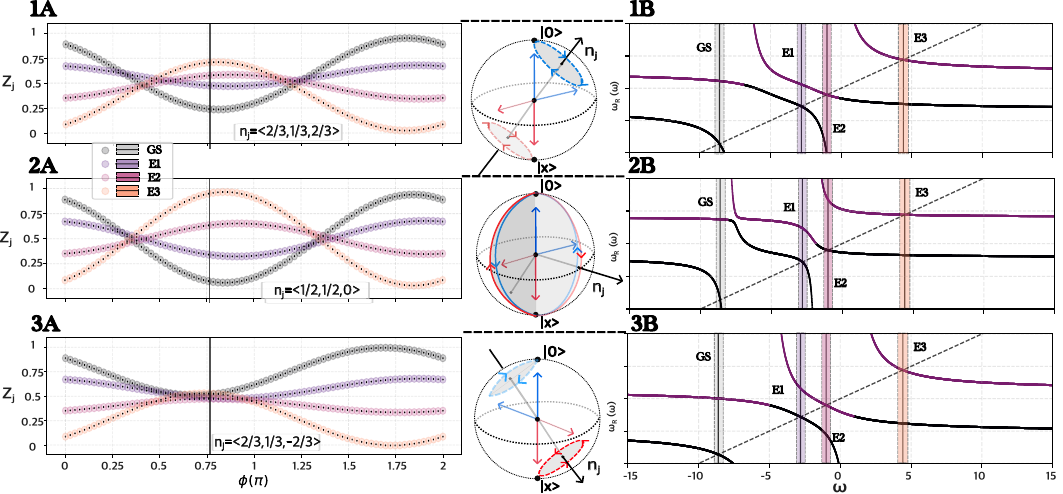}
\caption{ Renormalization and degree of separability under various bath states, generated via sequential unitary transformations given in Eq.~\ref{bath state and rotation }. \textbf{1-3A)} The degree of separability at three rotation axes $\mathbf{n}_j$ are defined via the corresponding Bloch spheres, which are depicted in the middle column. \textbf{1-3B} Interaction curves  $\omega_{\mathcal{R}}(\omega)$ at the rotation angle $\phi=0.76 \pi$. The variations in the frequency landscape near the fixed points are direct measures of the degree of separability depicted in \textbf{1-3A}. The sweep is configured with the following parameters:  $\omega_0=6$, $\omega_d=2+2\rm i$, $V_{00}=1.5$, $V_{0x}=J_{0x}=1$, $V_{xx}=0.5$. The variation near fixed points (black solid vertical lines) directly reflects the degree of separability of eigenstates of the ``universe''.      
    }
    \label{fig smooth curve}
\end{figure*}

\section{Numerical demonstration for the 2-site fermionic system}
\begin{figure*}[bt!]
    \centering
    \includegraphics{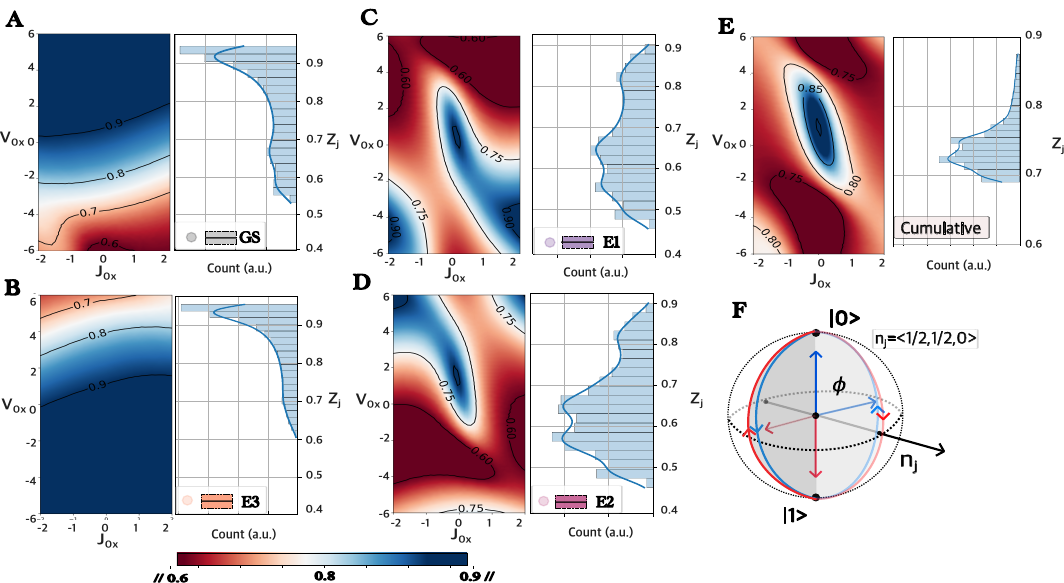}
    \caption{\textbf{A-D} (left) Maximum degree of separability of 4 individual eigenstates 
    on the $(J_{0x},V_{0x})$ plane. (right) The statistical distribution of {all} maximum degree of separability within the $(J_{0x},V_{0x})$ plane. \textbf{A, C} and \textbf{B, D} are connected by connecting $H\to -H$. \textbf{E} An {average} maximum degree of separability of 4 eigenstates. \textbf{F} The Bloch sphere representation of the unitary transformation, and the optimization is performed across the shaded area within the sphere. It is interesting to note that various density-density and exchange interaction parameters between site \textbf{0} and \textbf{x} generate separable eigenstates (colored in blue). In particular, the asymmetry of the maximum degree of separability in the second (third) excited state E1 (E2) suggests a qualitative difference between attractive $V_{0x},J_{0x}<0$ and repulsive $V_{0x},J_{0x}>0$ interactions. }
    \label{fig element heatmap}
\end{figure*}

To illustrate the concepts discussed above,  {let's consider a universe with 2 sites, labelled by \textbf{0} and \textbf{x}. We add two electrons with opposite spin $\sigma =\uparrow,\downarrow$. The Hilbert space of the universe is then given by: $\{\ket{0\uparrow0\downarrow}, \ket{0\uparrow x\downarrow}, \ket{x\uparrow0\downarrow}, \ket{x\uparrow x\downarrow}\}$. Although we use a fermionic example, the general formulation does not restrain the underlying statistics (fermionic or bosonic) since it is incorporated in the coupling kernel $\epsilon\boldsymbol{\mathcal{K}}$.} The 2-site Hamiltonian that we consider has the following form, 
\begin{equation}
\begin{aligned}
    H=&  \sum_{\sigma}-\frac{3}{4}\omega_0  c_{0,\sigma}^{\dag} c_{0,\sigma}+\frac{1}{4}\omega_0 c_{x,\sigma}^{\dag} c_{x,\sigma} + \frac{\omega_d}{2} c_{x,\sigma}^{\dag} c_{0,\sigma}+ c.c. \\
&+ \sum_{\bar{\sigma},\sigma}V_{00}n_{0,\sigma}n_{0,\Bar{\sigma}}+V_{0x}n_{0,\sigma}n_{x,\Bar{\sigma}}+V_{xx}n_{x,\sigma}n_{x,\Bar{\sigma}}\\
&+J_{0x}(c^\dag_{0,\downarrow}  c^\dag_{x,\uparrow}  c_{x,\downarrow}c_{0,\uparrow} +c^\dag_{x,\downarrow}  c^\dag_{0,\uparrow}  c_{0,\downarrow}c_{x,\uparrow} ),
\end{aligned}
\label{Eq. sys Hamiltonian}
\end{equation}
which is characterized by an energy scale $\omega_0$, the hopping strength $\mathrm{Re}\{\omega_d\}$, and the transition dipole moment strength $\mathrm{Im}\{\omega_d\}$. When $\mathrm{Im}\{\omega_d\} \equiv 0$, Eq.~\ref{Eq. sys Hamiltonian} gives a Hubbard-like model \cite{altland_simons_2010}, and when $\mathrm{Re}\{\omega_d\} \equiv 0$, it gives a Jaynes–Cummings-like model \cite{walls2008quantum}. The $\sigma,\bar{\sigma} \in \{\uparrow,\downarrow\},\{\downarrow,\uparrow\}$ labels the spins of electrons. The two-body terms, namely the density-density interaction $V$ and the exchange interaction term $J$, characterize the local and non-local electron-electron correlation effects. Let's select a fixed bath state $\ket{\text{bath, }j=0\uparrow}=\ket{0\uparrow}$ (spin-up electron at site 0). {This means that we pinned the degrees of freedom of the spin-up electron to site $\mathbf{0}$, while preserves the spin-down electron degrees of freedom. The pinned spin-up electron exert static interaction on the ``freely moved" spin-down electron}, and the corresponding static, rest space, and coupling Hamiltonian have a block form as given in the supplementary material. Using Eq.~\ref{self energy and rn block def}, we define the renormalized Hamiltonian, 
\begin{equation}
    \mathbf{H}^{{\mathcal{R}}}_{0\uparrow} (\omega_\mathcal{R}) =  \mathbf{H}^{\mathcal{S}}_{0\uparrow}+ \epsilon \boldsymbol{ M}(\omega), \hspace{0.3cm} \boldsymbol{ M}(\omega)=\mathbf{C}_{0x} \frac{1}{\omega-\mathbf{H}_{x\uparrow}} \mathbf{C}_{0x}^\dag,
\end{equation}
where renormalized eigenvalues and eigenstates are computed through the non-linear eigenvalue problem: 
\begin{equation}
\mathbf{H}^{{\mathcal{R}}}_{0\uparrow}(\omega) \ket{\boldsymbol{{\mathcal{R}}_{0\uparrow}}} = \omega_{{\mathcal{R}}}(\omega) \ket{\boldsymbol{ {\mathcal{R}}_{0\uparrow}}}.
\end{equation}
 As illustrated in Fig.~\ref{fig element heatmap} 1-3B), the intersections (vertical colored lines) between $\omega$ and $\omega_{{\mathcal{R}}}(\omega_\lambda)$ recover the universe's eigenvalues (represented by vertical black lines), labelled as GS, E1, E2, E3. Four $\omega_{\mathcal{R}}(\omega)$ interaction curves (eigenvalues of $\boldsymbol{ M}(\omega)$) smoothly connect the corresponding fixed points. Following Eq.~\ref{grad curve}, we numerically calculate the frequency variation near each fixed point, recovering the degree of separability $\boldsymbol{{Z}_j}$ in Eq.~\ref{equality}.

We have the freedom to choose any bath state, which can be generated from an initial bath state via a unitary transformation $\ket{\rm bath, l} = \mathbf{U}_{lj} \ket{\rm bath, j}$. This transformation corresponds to selecting a different measurement basis. While the system energy spectrum is preserved, i.e., $\mathbf{H} = \mathbf{U}_{lj} \mathbf{H} \mathbf{U}_{lj}^\dag$ under this transformation, the interaction kernel $\boldsymbol{\mathcal{K}}[\mathbf{U}_{lj}]$ now exhibits a functional dependence on the unitary operator, leading to a redistribution of the degree of separability $ \boldsymbol{{Z}}_l[\mathbf{U}_{lj},\omega_{\lambda}]$. For example, a separability coefficient that was initially spread out over a wide spectrum could become more narrowly peaked after the unitary transformation. By sweeping through different bath configurations, we can observe the variation of the degree of separability, as depicted in Fig.\ref{fig smooth curve}. 
 
To demonstrate the effects of bath states, we define a unitary operator that generates pairwise rotations with respect to a given axis $\mathbf{n}_j$ in the Bloch sphere, represented by:
\begin{equation}
   \mathbf{U}_{j}[\boldsymbol{\phi}_j]=\mathrm{Exp}(-i(\mathbf{n}_j\cdot \boldsymbol{\sigma})\frac{\phi}{2})= \mathrm{cos}(\frac{\phi}{2})\mathbf{I}-i(\mathbf{n}_j\cdot \boldsymbol{\sigma})\mathrm{sin}(\frac{\phi}{2}),
       \label{bath state and rotation }
\end{equation}
where $\phi$ is the rotation angle, and $\boldsymbol{\sigma}=(\sigma^1, \sigma^2, \sigma^3)$ represents Pauli matrices given in the supplementary material. Elements in the Bloch sphere generate unitary operators that transform the initial bath state $\ket{\text{bath, }j=0\uparrow}$ into a new bath state. As depicted in the middle column of Fig.~\ref{fig smooth curve},  two initial bath states and the reference axis are labeled as $\ket{0}$, $\ket{x}$, and $\mathbf{n}_j$, respectively. Each Bloch sphere structure represents a particular family of unitary transformations, labelled by the angle in Fig.~\ref{fig smooth curve} 1-3A). As we vary the rotation angle, the fixed solutions remain invariant, and depending on the rotation axis, the degree of separability oscillates and reflects the entropy lower bound in Eq.~\ref{lower bound equation} at given bath states. The ground state (\textbf{GS}) and the highest excited state (\textbf{E3}) have nearly unity degree of separability under some fixed bath states, while the two mid-excitation levels, \textbf{E1} and \textbf{E2}, exhibit entanglements under all bath states. {This can be understood intuitively by considering the concept of state symmetry, energy, and level repulsion. It is evident that the eigenstates with energy $E_1$ ($GS$) and $E_2$ ($E3$) have the same symmetry by considering the operation $H \to -H$. Since $E_1$ and $E_2$ are closer in magnitude, the level repulsion (avoid-crossing) contributes significantly to the state mixing, resulting in a non-product state like eigenstates \cite{Salem1975}. }

\begin{figure*}[bt!]
    \centering
    \includegraphics{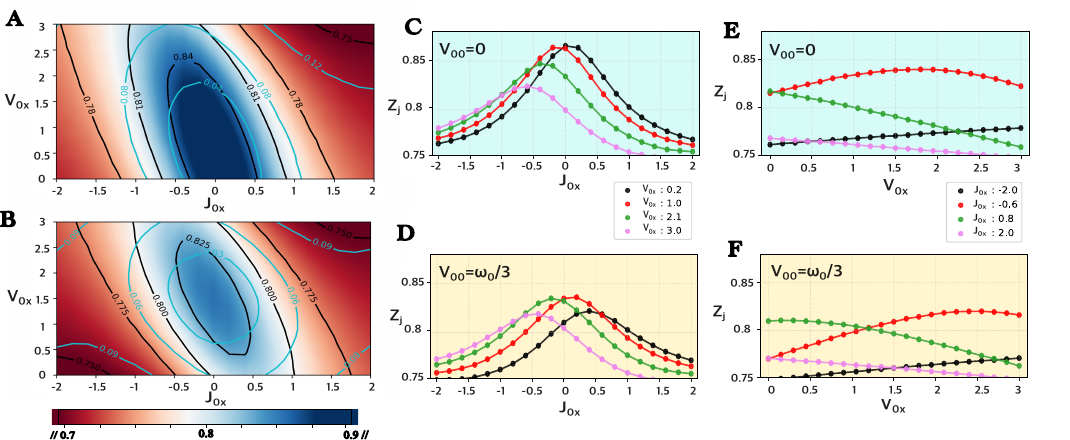}
     \caption{ \textbf{A,B)} The zoom-in maximum degree of separability heatmaps on the $(J_{0x},V_{0x})$ plane with two different on-site repulsion strength: \textbf{A)} $V_{00}=0$; \textbf{B)} $V_{00}=2$. The cyan contours label the standard deviation (Fig.S1) of the maximum degree of separability. \textbf{C,D} The maximum degree of separability landscape while varying $J_{0x}$. \textbf{E,F} The maximum degree of separability landscape while varying $V_{0x}$. The above calculations feature a fundamental gap of $\omega_0=6$, hopping (dipole) strength of $\omega_d=3+3\mathrm{i}$, and an on-site interaction strength of $V_{xx}=1$. }
    \label{fig different V00}
\end{figure*}

We further investigate the {maximum} degree of separability across {all} fixed bath states at a given set of interaction parameters. We perform numerical calculations to determine the maximum degree of separability at each point on the $(J_{0x},V_{0x})$ plane, and extract the counting statistics, as shown in Fig.\ref{fig element heatmap}. The maximum degree of separability heatmaps for the ground state (GS) in Fig.\ref{fig element heatmap} A and the highest eigenstate (E3) in Fig.\ref{fig element heatmap} B reveal eigenstates with swapped orders due to a $\pi$ rotation, a consequence of the simple Hamiltonian relation $H \to -H$. A similar argument applies to E1 in Fig.~\ref{fig element heatmap} C and E2 in Fig.~\ref{fig element heatmap} D. In Fig.~\ref{fig element heatmap}, we readily pinpoint the regime $(V_{0x} \approx [-1,2], J_{0x} \approx [-0.4,0.4])$ where {all} four eigenstates can be approximated as separable states. 

We then look at the averaged maximum degree of separability under two fixed $V_{00}$ values, as depicted in Fig.\ref{fig different V00} (A-B). An increase in the strength of $V_{00}$ reduces the average maximum degree of separability, and the variation with respect to the exchange interaction strength $J_{01}$ is illustrated in Fig.\ref{fig different V00} (C-D). On the other hand, as shown in Fig.~\ref{fig different V00} (E-F), the variation with respect to the Coulomb interaction strength $V_{01}$ has a weak dependence.

This formulation provides an alternative method for systematically identifying regions in the parameter space where the corresponding eigenstates can be approximated as separable states. Such an approach can offer valuable insights into the interplay between interaction parameters and the effects of state hybridization. For example, as suggested in Fig.~\ref{fig element heatmap}, having both attractive (repulsive) density-density interaction and repulsive exchange interaction generates near separable first (second) excited state. 

\section{Weak interaction limit }

  While Eq.~\ref{equality} is exact, it is challenging to evaluate analytically and can be significantly simplified under a few physically motivated assumptions in the weak interaction limit: 
\begin{itemize}
    \item We assume that the degree of similarity $\mathbf{z_j}( \omega-\omega')$ has a value close to unity around the corresponding renormalized Hamiltonian eigenvalue within the cutoff energy $\Omega$, such that $1  \gtrsim \mathbf{z_{j}}( \Omega) \gg 0$. Under this assumption, as shown in the supplementary material, the degree of separability in Eq.~\ref{equality} is dominated by the diagonal term near a fixed point.  

\item For weakly interacting many-body problems ($\epsilon \approx 0)$, eigen information of the renormalized Hamiltonian is related to the static Hamiltonian via perturbation theory. The first-order non-degenerate perturbation expansion of a renormalized Hamiltonian eigenvalue near a static Hamiltonian eigenvalue $\omega_{\mathcal{S}}$ has an implicit form, 
\begin{equation}
     \omega_{\lambda} \approx \omega_{\mathcal{S}} + \epsilon \boldsymbol{M_{\mathcal{S}\mathcal{S}}}(\omega_{\lambda}) +O(\epsilon^2),
\end{equation}
where we define $  \boldsymbol{M_{\mathcal{S}\mathcal{S}}}(\omega_{\lambda})=\braket{\boldsymbol{\mathcal{S}_j}|\boldsymbol{ M}(\omega_{\lambda}) |\boldsymbol{S_j}} $. 
Up to the first order correction, a renormalized eigenstate $\ket{\boldsymbol{\mathcal{R}_j}}$ can be approximated as the following, 
\begin{equation}
\ket{\boldsymbol{\mathcal{R}_j}} 
 \approx \ket{\boldsymbol{\mathcal{S}_j}} +  \epsilon \sum_{\mathcal{S}' \neq \mathcal{S}} \frac{\boldsymbol{M_{\mathcal{S}\mathcal{S}'}}(\omega_{\lambda}) }{ \omega_{\mathcal{S}} -\omega_{\mathcal{S}'}}\ket{\boldsymbol{\mathcal{S}'_j}}+O(\epsilon^2),
 \label{scattering}
\end{equation}
and the degree of similarity is inferred as $\mathbf{z_j}( \omega_{\mathcal{S}\lambda}) \approx 1+{\mathcal{O}}(\epsilon^2)$, as shown in the supplementary material. {Eq.~\ref{scattering} is in direct analogy to the Lippmann–Schwinger equation in scattering theory \cite{PhysRev.79.469}}.

\item We also make the assumption that the diagonal element dominates the coupling kernel near a fixed point $\boldsymbol{M_{\mathcal{S} \mathcal{S}}}(\omega_{\lambda}) \gg \boldsymbol{M_{\mathcal{S} \mathcal{S'}}}(\omega_{\lambda})$, and the lowest order of the frequency variation is linear, 
    \begin{equation}
   -\left. \frac{\partial \omega_{{\mathcal{R}}}(\omega) }{\partial \omega}\right|_{\omega_{\lambda }}   \approx \epsilon \sum^{\rm All}_{\mathcal{S}', \mathcal{S}''} \frac{ \boldsymbol{M_{\mathcal{S} \mathcal{S}'}} \boldsymbol{M_{\mathcal{S}'' \mathcal{S}}}}{{\boldsymbol{\mathcal{K}}_{\mathcal{S}' \mathcal{S}''}} } \approx \epsilon 
 \frac{ (\boldsymbol{M_{\mathcal{S} \mathcal{S}}}(\omega_{\lambda}))^2  }{{\boldsymbol{\mathcal{K}}_{\mathcal{S} \mathcal{S}}} },
 \label{frequency variation}
 \end{equation}
where the kernel element is defined in Eq.~\ref{kernel}. This is motivated by the famous random phase approximation (RPA) in many-body perturbation theory. RPA suggests that the Coulomb interaction between particles are ``screened'' by the vacuum bubbles created from the density fluctuation of the dielectric medium \cite{doi:10.1080/00268976.2011.614282,martin_reining_ceperley_2016}. The dominance of the RPA process can be attributed to its larger phase space volume \cite{altland_simons_2010}. Similarly, we assume that the diagonal part of the interaction kernel near a fixed point captures greater "configurational weights" than the off-diagonal counterparts. As a consequence, the sum rule in Eq.~\ref{sum w} are preserved within the cutoff frequency $\Omega$. 
 
\end{itemize}

Under the above assumptions, to the lowest order, the weight factor near a fixed point can be approximated by a standard Lorentzian,  
\begin{equation}
\boldsymbol{\mathcal{\Tilde{W}}_j} ( \omega_{\mathcal{S}\lambda})\approx\left(\frac{1}{\pi \sqrt{\epsilon{\boldsymbol{\mathcal{K}}_{\mathcal{S} \mathcal{S}}}  }} \right) \left({1+ 
 \frac{ (\omega_{\lambda}-\omega_{\mathcal{S}  })^2  }{\epsilon{\boldsymbol{\mathcal{K}}_{\mathcal{S} \mathcal{S}}} }}\right)^{-1}, 
    \label{weight factor approx}
\end{equation}
where we rescale the original weight factor by an overall normalization constant,
\begin{equation}
    \boldsymbol{\mathcal{\Tilde{W}}_j} ( \omega_{\mathcal{S}\lambda})=\left({\pi \sqrt{\epsilon{\boldsymbol{\mathcal{K}}_{\mathcal{S} \mathcal{S}}}  }} \right)\boldsymbol{\mathcal{{W}}_j} ( \omega_{\mathcal{S}\lambda}) ,
\end{equation}
and use the relation $\omega_{\mathcal{S}\lambda}\approx \epsilon\boldsymbol{M_{\mathcal{S} \mathcal{S}}}(\omega_{\lambda})$. It's worth noting that the rescaled weight factor is {discrete} and doesn't have {any} free variables, and it is only defined at universe's eigenvalues that are within the cutoff frequency $\Omega $. 

\begin{figure*}[bt!]
    \centering
    \includegraphics{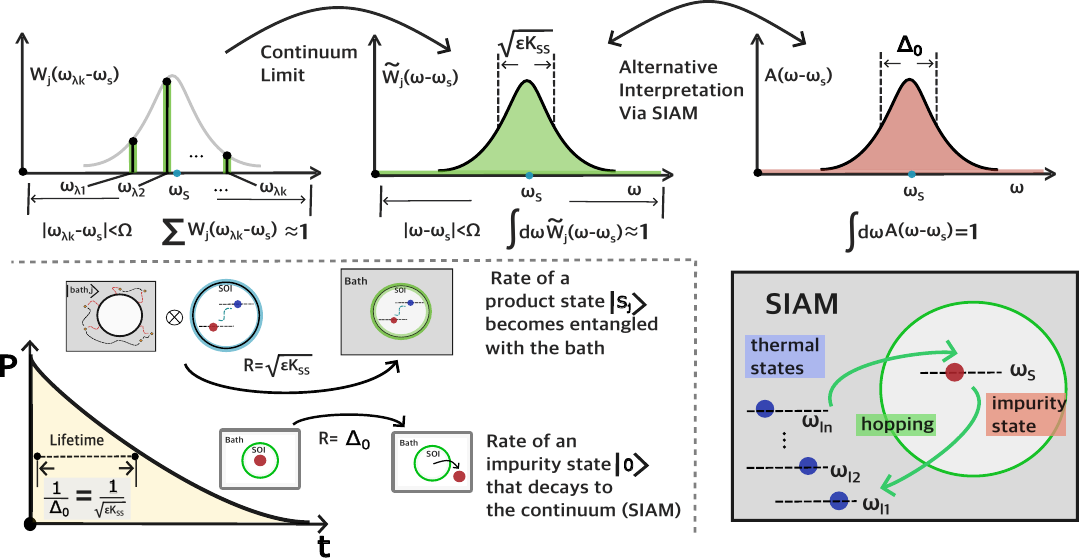}
     \caption{As illustrated on the top left, for the universe with finite Hilbert space dimension, it has discrete eigenvalues $\omega_{\lambda1},\omega_{\lambda2},\dots,\omega_{\lambda k}$, within the cutoff frequency $\Omega$, and the weight factor obeys a discrete sum rule. By taking the continuum limit around a given static eigenvalue $\omega_{\mathcal{S}}$, 
     \textbf{Top, middle}, the rescaled weight factor preserves the sum rule, and the distribution of the rescaled weight factor becomes a Lorentzian with a broadening proportional to the coupling kernel $\epsilon \kappa_{\mathcal{SS}}$. We then relate the weight factor distribution to the spectral function in the SIAM model, where the rate of an initially prepared product state that becomes entangled to the bath can be understood as the rate of an initially prepared impurity state decaying to the continuum. $\mathbf{P}$ on the bottom left represents the probability of having a product state (in our formalism) or of having an impurity state (in the case of the SIAM) over time $t$, highlighting the correspondence. 
 }
    \label{branch cut}
\end{figure*}

A linear increase in bath size leads the rest space dimension to grow exponentially, e.g. $\Lambda \sim \mathrm{exp}[\mathcal{D}_{\rm bath}]$. In the limit $\Lambda \to \infty$, $\omega_{\lambda}$ form branch cuts on the real axis, and as a result, universe's eigenvalues becomes a continuous variable. Within the cutoff frequency $|\omega_\lambda -\omega_{\mathcal{S}}|<\Omega$, the sum rule has an integral form in the continuum limit $\Lambda \to \infty$,  $1=\lim_{\Lambda \to \infty} \sum_\lambda \boldsymbol{\mathcal{W}_j} ( \omega_{\mathcal{S}\lambda})$,
\begin{equation}
\begin{aligned}
     1 &  \approx \sum_\lambda \pi \sqrt{\epsilon{\boldsymbol{\mathcal{K}}_{{\mathcal{S}} {\mathcal{S}}}}  }  \left(\frac{1}{\pi \sqrt{\epsilon{\boldsymbol{\mathcal{K}}_{{\mathcal{S}} {\mathcal{S}}}}  }} \right) \left({1+ 
 \frac{ (\omega_{\lambda}-\omega_{\mathcal{S}})^2  }{\epsilon{\boldsymbol{\mathcal{K}}_{{\mathcal{S}} {\mathcal{S}}}} }}\right)^{-1} \\
 &=\sum_\lambda \pi \sqrt{\epsilon{\boldsymbol{\mathcal{K}}_{{\mathcal{S}} {\mathcal{S}}}}  } \boldsymbol{\mathcal{\Tilde{W}}_j} ( \omega_{\mathcal{S}\lambda})\\
 &=\int \mathrm{d}\omega \rho (\omega)\pi \sqrt{\epsilon{\boldsymbol{\mathcal{K}}_{{\mathcal{S}} {\mathcal{S}}}}  } \boldsymbol{\mathcal{\Tilde{W}}_j} ( \omega)=\int \mathrm{d}\omega \boldsymbol{\mathcal{\Tilde{W}}_j} ( \omega-\omega_{\mathcal{S}}),
\end{aligned}
\label{sum rule continuum}
\end{equation}
where $\rho(\omega)=\left(\pi \sqrt{\epsilon{\boldsymbol{\mathcal{K}}_{{\mathcal{S}} {\mathcal{S}}}}  } \right)^{-1}$ is a constant that enforces the normalization condition. In the non-interacting limit, we recover the Dirac delta function $ \lim_{ \epsilon \to 0}  \boldsymbol{\mathcal{\Tilde{W}}_j} ( \omega-\omega_{\mathcal{S}})=\delta ({\omega-\omega_{\mathcal{S} } })$. 

\section{Correspondence between quasiparticle dissipation and quantum information decay}
  
  Let's consider a static Hamiltonian eigenstate (bare particle), $\ket{\boldsymbol{\mathcal{S}_j}}$, with energy $\omega_{\mathcal{S}}$. As we gradually turn on the perturbation $\epsilon$, the SOI and the bath become entangled, leading to {an energy continuum $\omega_\lambda$}, emerging near the static Hamiltonian eigenvalue $\omega_{\mathcal{S}}$. {Eq.~\ref{weight factor approx} suggests that to first order in $\epsilon$, the probability density of observing the product state $\ket{\boldsymbol{\mathcal{S}_j}}$ with energy $\omega$ in the universe follows a standard Lorentzian \cite{altland_simons_2010,martin_reining_ceperley_2016} with a broadening of $\sqrt{\epsilon{\boldsymbol{\mathcal{K}}_{\mathcal{S} \mathcal{S}}}}$. }

This formulation provides an alternative method for systematically identifying regions in the parameter space where the corresponding eigenstates can be approximated as separable states. Such an approach can offer valuable insights into the interplay between interaction parameters and the effects of state hybridization. For example, as suggested in Fig.~\ref{fig element heatmap}, having both attractive (repulsive) density-density interaction and repulsive exchange interaction generates near separable first (second) excited state.

  Within the cutoff frequency, Eq.~\ref{weight factor approx} can then be expressed as the imaginary part of a complex-valued function:
\begin{equation}
    \boldsymbol{G} ({\omega})\equiv \frac{1}{{\omega}-\omega_{\mathcal{S}}-i\epsilon{\boldsymbol{\mathcal{K}}_{\mathcal{S} \mathcal{S}}}}, \hspace{0.4cm} |\omega - \omega_{\mathcal{S}}|<\Omega,
    \label{OQS Greens}
\end{equation}
analogous to the single-particle Green's function in an impurity Anderson model \cite{PhysRev.124.41}. Specifically, let's consider a single-impurity Anderson model (SIAM), where an impurity site is coupled to a thermal bath. The model Hamiltonian has a second quantization form,
\begin{equation}
\begin{aligned}
\mathbf{H}&=\mathbf{H}_{\rm local}+\mathbf{H}_{\rm couple}+\mathbf{H}_{\rm bath}\\
        &=  \omega_{\mathcal{S}} c^\dag_{\mathcal{S}} c_{\mathcal{S}} +\sum_{l}^{\mathcal{D}_{\rm bath}} \left(t_{ {\mathcal{S}} l} c^\dag_{\mathcal{S}} a_l+ t^*_{ {\mathcal{S}} l}  a^\dag_l c_{\mathcal{S}}+ \omega_l a^\dag_l a_l\right) ,
\end{aligned}
\label{impurity H}
\end{equation}
where $c^\dag_{\mathcal{S}} (c_{\mathcal{S}})$ and $a^\dag_l(a_l)$ are the creation(annihilation) operators of an impurity state and a thermal state, respectively. The propagation of the impurity state is captured by the corresponding impurity Green's function, 
\begin{equation}    \boldsymbol{\mathcal{G}}_{\mathcal{R}}(\omega)=\lim_{\eta \to 0^+}\frac{1}{\omega-\omega_{\mathcal{S}}-\boldsymbol{\Delta}_{\mathcal{R}}(\omega + i\eta)},
\label{embedding Greens}
\end{equation} 
where the retarded hybridization function $\boldsymbol{\Delta}_{\mathcal{R}}(\omega+i\eta)$ reflects the effective {causal} interaction between an impurity state and thermal states. In the continuum limit, the hybridization function transforms into a function of the density of states $D(\omega)$ and a continuous coupling function $t_{{\mathcal{R}}}(\omega)$, as shown in the supplementary material. If both the density of states and the coupling function are approximated as constants, the hybridization function simplifies to an imaginary constant, $\boldsymbol{\Delta}_{{\mathcal{R}}}(\omega+ i\eta)=-i|\Delta_0|$, representing the rate for a particle initially in the impurity state to escape into the bath, and is defined as $|\Delta_0|= D_0 (t_0)^2$ \cite{martin_reining_ceperley_2016}.

The many-body spectral function is defined as the imaginary part of the impurity Green's function, $\mathbf{{A}}(\omega)=-\mathrm{Im}\{\boldsymbol{\mathcal{G}}_{\mathcal{R}}(\omega )/\pi\}$, and it takes the following form,
\begin{equation}
    \mathbf{{A}}(\omega-\omega_{\mathcal{S}}) = \frac{1}{\pi} \frac{|\Delta_0|}{(\omega-\omega_{\mathcal{S}})^2 + |\Delta_0|^2  }.
        \label{spectral function embedding}
\end{equation} 
It is evident that the impurity spectral function defined in Eq.~\ref{spectral function embedding} shares an identical standard Lorentzian dependence with the rescaled weight factor described in Eq.~\ref{weight factor approx} by substituting the following:  
\begin{equation}
|\Delta_0| \equiv \sqrt{\epsilon\boldsymbol{\mathcal{K}}_{\mathcal{S} \mathcal{S}}} , \hspace{0.4 cm} \mathbf{{A}} ( \omega-\omega_\mathcal{S}) \equiv \boldsymbol{\mathcal{\Tilde{W}}_j} ( \omega-\omega_\mathcal{S}).
 \label{maps}
\end{equation}
It immediately suggests that an alternative interpretation: the rate $|\Delta_0|$ for a particle initially in the impurity state to escape into the continuum can be regarded as the rate $\sqrt{\epsilon\boldsymbol{\mathcal{K}}_{\mathcal{S} \mathcal{S}}}$ {at which a static Hamiltonian eigenstate $\ket{\boldsymbol{\mathcal{S}_j}}$ becomes entangled to the bath degrees of freedom under many-body interaction}. In the non-interacting limit, $\epsilon \to 0$, the rate becomes zero, i.e. the initial state remains isolated from the bath.

The time-domain propagation in response to the bath coupling is given by the inverse Fourier transform of Eq.~\ref{OQS Greens}: 
 \begin{equation}
\boldsymbol{{G}}_{\mathcal{R}}(t)=\mathcal{F}^{-1}\left[\boldsymbol{{G}}_{\mathcal{R}}(\omega)\right]=-i \Theta(t) \mathrm{Exp}\left(-i\omega_{\mathcal{S}} t -|\Delta_0|  t\right), 
 \end{equation}
 where $\Theta(t)$ is the step function. The time evolution has two parts, an oscillating and a decay part, reflecting the coherence evolution within the SOI and decoherence due to the bath coupling, respectively. 
 
The decoherence of a universe eigenstate around a static eigenstate can be further interpreted as the energy gain/loss during a fictitious particle exchange event. Let's consider an {effective} two-level {non-Hermitian} Hamiltonian, which describes the transport of a fictitious particle into and out of the bath. The fictitious particle has an intrinsic energy of $m_\phi = \omega_{\mathcal{S}}$, and we use the labels $\ket{0}$ and $\ket{1}$ to indicate the particle's location, whether it's in the SOI or bath. In addition, the fictitious particle exchange will add or remove an energy of $\pm |\Delta_0|$ from the respective systems. The effective Hamiltonian takes a matrix form,
\begin{equation}
   H_{\rm eff}= \begin{bmatrix}
        \omega_{\mathcal{S}} & |\Delta_0|\\
        -|\Delta_0| &  \omega_{\mathcal{S}}\\
    \end{bmatrix},
    \label{2 level Hamiltonian}
\end{equation}
which has two {complex} eigenvalues $\lambda_{\rm eff}= \omega_{\mathcal{S}} \pm i |\Delta_0|$, {reflecting the particle transfer in and out of the bath, respectively}. The corresponding Green's function can be evidently written as the expectation value of single-particle time evolution,
 \begin{equation}
\boldsymbol{\mathcal{G}}_{\mathcal{R}}(t)=-i \Theta(t) \braket{\Omega|\psi_0 e^{-iH_{\rm eff}t} \psi_0^\dag |\Omega},
\label{greens function exchange}
 \end{equation}
where $\ket{G}$ is the zero-particle ground state with energy $E_{G}\equiv0$, and $\psi_0^\dag/\psi_0$ creates/annihilates the fictitious particle.

In summary, as suggested in Eq.\ref{maps}, when near a fixed point, the impurity spectral function in SIAM can be interpreted as the rescaled weight factor when the SOI is weakly entangled with the bath. {The hybridization function $|\Delta_0|$ in SIAM can be related to the coupling kernel $\sqrt{\epsilon\boldsymbol{\mathcal{K}}_{\mathcal{S} \mathcal{S}}}$, and the loss of the degree of separability (lower-bound of entropy gain) can be modeled as the decay of a fictitious particle from the SOI to the bath, as given in Eq.\ref{greens function exchange}.} Let's go back to the question we asked earlier that involves two interacting excitons in a box. Although we can always recover universe's eigenvalues using a renormalized Hamiltonian, we must have a near-unity degree of separability $\boldsymbol{{Z}_j}(\omega-\omega_{\mathcal{S}})$ to faithfully represent the their interaction as classical density-density interaction. When their coupling is weak, we can approximate the degree of separability by an equivalent many-body impurity problem.

\section{ Discussion and Conclusion}
In this paper, we propose a novel approach to quantify the quantum information lost to the bath and its connection to the quasiparticle spectral function. We conclude that the degree of separability of an energy eigenstate, as defined in Eq.\ref{equality}, is an essential measure that establishes the lower bound of SOI-bath entanglement entropy, as discussed numerically in a 2-level model. While the entanglement entropy between system and bath remains invariant under unitary transformation, in practice, we may want to use a product state to best approximated the universe's eigenstate. For near unity degree of separability, at a given energy, we can model the SOI-bath interaction as a classical external potential. Furthermore, we demonstrate that when the coupling is weak, the weight factor, defined in Eq.\ref{weight factor approx}, follows a standard Lorentzian with a broadening of $\sqrt{\epsilon{\boldsymbol{\mathcal{K}}_{\mathcal{S} \mathcal{S}}}}$. We then establish a connection between the entanglement measure and the energy exchange processes that occur in an effective non-Hermitian interaction, as described in Eq.~\ref{greens function exchange}. 

Although we only presented the simple case where the decoherence process of a many-body eigenstate is mapped to a single fictitious quasiparticle propagation, this mapping might be extendable beyond the weakly interacting limit and from bipartite system to multipartite system. The multipartite entanglement can be computationally observed via concentratable entanglements \cite{PhysRevLett.127.140501, PhysRevA.106.042411}, and its relation to quasiparticle propagation can be further examined. Our findings open a new direction for bridging concepts between many-body physics and quantum information science.

\section{Acknowledgement}
We acknowledge support from NSF Award No. 2427169, 2137740 and Q-AMASE-i, through Grant No. DMR-1906325, and from NWO Quantum Software Consortium (Grant No. 024.003.037).  Furthermore, the authors thank Sagar Vijay and Vojtech Vlcek for their insightful discussions and suggestions.

\appendix 
\bibliographystyle{unsrt} 
 \widetext
\section{Appendix}

\textbf{\sffamily Fixed point solution  } The fixed point equation $\omega_{\lambda}=\omega_{{\mathcal{R}}}(\omega_\lambda)$ can be expanded at linear order around a fixed point, 
\begin{equation}
\begin{aligned}
        &(\omega_{\lambda}-\omega_{\mathcal{S}})+\left( \omega_{\mathcal{S}}-\omega_{{\mathcal{R}}}(\omega_{\mathcal{S}})- \left(\left. \frac{\partial \omega_{{\mathcal{R}}}(\omega)}{\partial \omega}\right|_{\omega_{\lambda}}\right)(\omega_{\lambda}-\omega_{\mathcal{S}}) \right)=0, \hspace{0.5cm} \text{with solution, }  (\omega_{\lambda}-\omega_{\mathcal{S}})=\boldsymbol{{Z}}_j \left(\omega_{\mathcal{S}}-\omega_{{\mathcal{R}}}(\omega_{\mathcal{S}})\right),
\end{aligned}
\label{fixed point equation step}
\end{equation}
and $\boldsymbol{{Z}}_{j}$ is defined in Eq.~\ref{grad curve}.

\textbf{\sffamily Renormalized Hamiltonian} 
The relation in Eq.~\ref{first equations} leads to,  
\begin{equation}
     |b^j_\lambda|^2   =\epsilon|a^j_\lambda|^2\bra{\boldsymbol{{\mathcal{R}}_j}}\mathbf{C}\left(\frac{1}{\omega_{\lambda}-\mathbf{H}_{R|j}} \right)  \left( \frac{1}{\omega_{\lambda}-\mathbf{H}_{R|j}} \mathbf{C}^\dag  \right)\ket{\boldsymbol{{\mathcal{R}}_j}}.
     \label{Left equation Z factor}
\end{equation}
Eq.~\ref{first equations} also gives, 
\begin{equation}
\ket{\boldsymbol{{\mathcal{R}}_j}}=\epsilon\frac{1}{\omega_{\lambda}-\mathbf{H}^{\mathcal{S}}_j} \mathbf{C}\frac{1}{\omega_{\lambda}-\mathbf{H}_{R|j}} \mathbf{C}^\dag 
\ket{\boldsymbol{{\mathcal{R}}_j}},  \hspace{0.4cm}                \epsilon \mathbf{C}\frac{1}{(\omega_{\lambda}-\mathbf{H}_{R|j})} \mathbf{C^\dag} \ket{\boldsymbol{{\mathcal{R}}_j}}  =     (\omega_{\lambda}-   \mathbf{H}^{\mathcal{S}}_j )\ket{\boldsymbol{{\mathcal{R}}_j}}. \\
\end{equation}

\textbf{\sffamily Lower bound of the Von Neumann entropy } 
The bipartite splitting of the universe defines the reduced density matrix at a given eigenstate $\ket{\boldsymbol{\lambda}}$,
\begin{equation}
 \boldsymbol{\rho_{\rm SOI}} =\mathrm{Tr_{bath}}\left(\ket{\boldsymbol{\lambda}}\bra{\boldsymbol{\lambda}}\right)=\sum_j \braket{\rm bath,j \ket{\boldsymbol{\lambda}}\bra{\boldsymbol{\lambda}} \rm bath,j} =\sum_{j }^{\rm All}  \boldsymbol{{Z}_j}(\omega_{\mathcal{R}}=\omega_{\mathcal{\lambda}}) \ket{\boldsymbol{\Tilde{{\mathcal{R}}_j}}}  \bra{\boldsymbol{\Tilde{{\mathcal{R}}_j}}} , 
\end{equation}
{where $j$ labels independent basis vectors that span the N-dimensional Hilbert space with and,} 
\begin{equation}
  \ket{\boldsymbol{\Tilde{{\mathcal{R}}_j}}} =\braket{\rm bath,j |\boldsymbol{{\mathcal{R}}_j}},\hspace{0.4cm}  \sum_{j }^{\rm All}  \boldsymbol{{Z}_j}(\omega_{\mathcal{R}})=1, 
\end{equation}
with the Von Neumann entropy,
\begin{equation}
   \mathcal{E}(\omega_\lambda) = \sum_{j }^{\rm All}  \boldsymbol{{Z}}_j (\omega_{\mathcal{R}})\mathrm{Log}(\boldsymbol{{Z}}_j(\omega_{\mathcal{R}})).
\end{equation}
{Let's consider the case where we have selected a particular bath state $\ket{\rm bath,j}$ and obtain the corresponding eigenstate $\ket{\boldsymbol{{{\mathcal{R}}_j}}}$ from the renormalized Hamiltonian at the energy $\omega_\lambda$.} We constructed a density matrix $\boldsymbol{\rho_{\rm B}}$ at the fixed point $\omega_\lambda=\omega_{\mathcal{R}}$, 
\begin{equation}
     \boldsymbol{\rho_{\rm B}} =  \boldsymbol{{Z}_j}\ket{\boldsymbol{\Tilde{{\mathcal{R}}_j}}} \bra{\boldsymbol{\Tilde{{\mathcal{R}}_j}}} +\boldsymbol{{Z}_{j'}}\ket{\boldsymbol{\Tilde{{\mathcal{R}}_{j'}}}}  \ket{\boldsymbol{\Tilde{{\mathcal{R}}_{j'}}}} , 
\end{equation}
such that $\boldsymbol{{Z}_j}+\boldsymbol{{Z}_{j'}}=1$ with entropy, 
\begin{equation}
       B(\boldsymbol{{Z}_j})= -\boldsymbol{{Z}_j} \mathrm{Log}[\boldsymbol{{Z}_j} ]-(1-\boldsymbol{{Z}_j})  \mathrm{Log}[1-\boldsymbol{{Z}_j}].
\end{equation}
{We then demonstrate that $\boldsymbol{\rho_{\rm B}}$ has lowest entropy while fixing $\boldsymbol{{Z}_j}$.} Let's split the second part of the $\boldsymbol{\rho_{\rm B}}$ by a positively-defined leak parameter $\epsilon  \leq 1$, in Eq.~\ref{reduced density matrix LB} such that,
\begin{equation}
     \boldsymbol{\rho_{\rm B'}} =  \boldsymbol{{Z}_j} \ket{\boldsymbol{\Tilde{{\mathcal{R}}_{j}}}}  \ket{\boldsymbol{\Tilde{{\mathcal{R}}_{j}}}}+ 
 \epsilon \ket{\boldsymbol{\Tilde{{\mathcal{R}}_{j''}}}}  \ket{\boldsymbol{\Tilde{{\mathcal{R}}_{j''}}}}+\boldsymbol{{Z}_{j'}}\ket{\boldsymbol{\Tilde{{\mathcal{R}}_{j'}}}}  \ket{\boldsymbol{\Tilde{{\mathcal{R}}_{j'}}}}, 
     \label{reduced density matrix LB}
\end{equation}
with $ \epsilon+\boldsymbol{{Z}_j}+\boldsymbol{{Z}_{j'}}=1$ and $\epsilon \leq 1-\boldsymbol{{Z}_j}$. To the first order, the entropy is given by,
\begin{equation}
\begin{aligned}
        &    B'(\boldsymbol{{Z}_j},\epsilon)= -\boldsymbol{{Z}_j} \mathrm{Log}[\boldsymbol{{Z}_j} ]-\epsilon  \mathrm{Log}[\epsilon]-(1-\boldsymbol{{Z}_j}-\epsilon )\mathrm{Log}[1-\boldsymbol{{Z}_j}-\epsilon  ]\\
           & =B(\boldsymbol{{Z}_j})+\epsilon \left(1+\mathrm{Log}[1-\boldsymbol{{Z}_j}]-\mathrm{Log}\epsilon\right)+\mathcal{O}(\epsilon^2). 
\end{aligned}
\label{equation for lower bound}
\end{equation}
{where the last line, we use monotonic property of logarithmic function, i.e. $\mathrm{Log}(1-\boldsymbol{{Z}_j})-\mathrm{Log}(\epsilon) \geq 0$. Therefore, we have $B'(\boldsymbol{{Z}_j},\epsilon) \geq B(\boldsymbol{{Z}_j})$ for any $\epsilon \leq 1$. The result is not surprising: additional terms in the reduced density matrix will result in an increase in entropy {cite}. }

\textbf{\sffamily Interaction curve gradient} In particular, there exists a family of eigenstates $\ket{\boldsymbol{{\mathcal{R}}_j}}$ and eigenvalues $\omega_{{\mathcal{R}}}(\omega_\lambda)$ that satisfies the following equation,
\begin{equation}
\mathbf{H}^{{\mathcal{R}}}_{j} (\omega) \ket{\boldsymbol{{\mathcal{R}}_j}}=\omega_{{\mathcal{R}}}(\omega)\ket{\boldsymbol{{\mathcal{R}}_j}},
\label{eq:non_lin_eival}
\end{equation}
where $\omega_{{\mathcal{R}}}(\omega)$ defines an {interaction curve} that {smoothly} connects the fix point, i.e. $\omega_{\lambda}=\omega_{{\mathcal{R}}}(\omega_\lambda)$. Although those frequencies in Eq.~\ref{eq:non_lin_eival} that do {not} satisfy the fixed point equation have no direct physical meaning, near each fixed point, they are necessary for the gradient calculations. The first order derivative of the interaction curve has the form,
\begin{equation}
\begin{aligned}
         \left. \frac{\partial \omega_{{\mathcal{R}}}(\omega) }{\partial \omega}\right|_{\omega_{\lambda }} & =  \epsilon \bra{\boldsymbol{{\mathcal{R}}_j}}\mathbf{C} \frac{\partial}{\partial \omega} \left. \left(  \frac{1}{\omega- \mathbf{H}_{R|j}}\right)\right|_{\omega_{\lambda }} \mathbf{C}^\dag \ket{\boldsymbol{{\mathcal{R}}_j}}=-\epsilon\bra{\boldsymbol{{\mathcal{R}}_j}} \mathbf{C}  \left. \left(  \frac{1}{\omega_{\lambda}-\mathbf{H}_{R|j}}\right)\left(  \frac{1}{\omega_{\lambda}-\mathbf{H}_{R|j}}\right)\right. \mathbf{C}^\dag \ket{\boldsymbol{{\mathcal{R}}_j}}
          = - \frac{|b^j_\lambda|^2}{|a^j_\lambda|^2}. \\
\end{aligned}
\label{fixed point variation}
\end{equation}

\textbf{\sffamily Sum rule} the sum rule of the weight factor can be shown,
\begin{equation}
 1=  \braket{\boldsymbol{{\mathcal{R}}_j}|\boldsymbol{{\mathcal{R}}_j}} = \sum_{ \mathcal{S}} \braket{\boldsymbol{{\mathcal{R}}_j}|\boldsymbol{{\mathcal{S}}_j}}\braket{\boldsymbol{{\mathcal{S}}_j}|\boldsymbol{{\mathcal{R}}_j}}=\sum _{  \mathcal{S}} \mathbf{z_j} ( \omega_{\mathcal{S}\lambda}),
 \label{sum z}
\end{equation}
\begin{equation}
 1=  \braket{\boldsymbol{{\mathcal{S}}_j}|\boldsymbol{{\mathcal{S}}_j}} = \sum_{  \mathbf{\lambda}} \braket{\boldsymbol{{\mathcal{S}}_j}|\boldsymbol{\lambda}}\braket{\boldsymbol{\lambda}|\boldsymbol{{\mathcal{S}}_j}}=\sum_{ \mathbf{\lambda}} \boldsymbol{\mathcal{W}_j} ( \omega_\lambda-\omega_{\mathcal{S}}),
 \label{sum w}
\end{equation}
where we use the resolution of identity within the projected and universe's Hilbert space, respectively.

\textbf{\sffamily Singular value decomposition and interaction kernel } Let's recall the singular value decomposition (SVD), $\mathbf{ C= U(\Gamma)V^\dag}$, where $\mathbf{\Gamma}$ is a rectangular diagonal matrix, $ \mathbf{ U}$ and $\mathbf{ V}$ are rank $\mathcal{D}_{\rm sys}$ and $\Lambda-\mathcal{D}_{\rm sys}$ unitary square matrices, respectively. The product has an form $\mathbf{C}^\dag \mathbf{C}=\mathbf{V}(\Gamma^\dag\Gamma)\mathbf{V}^\dag$, and we assume there exist an interaction kernel, $\boldsymbol{\mathcal{K}} $ that is proportional to the coupling strength, so that $\mathbf{C^\dag ({1}/{\boldsymbol{\mathcal{K}}})C=I}$.

\textbf{\sffamily General matrix properties and weight factor expression} the rectangular coupling matrix can be decomposed via singular value decomposition (SVD),  
\begin{equation}
  \mathbf{ C= U(\Gamma)V^\dag, \hspace{0.5cm} C^\dag=V (\Gamma^\dag) U^\dag}, 
  \label{SVD}
\end{equation}
where $\boldsymbol{\Gamma}$ is a singular matrix. We define the interaction kernel, $\boldsymbol{\mathcal{K}} \equiv \mathbf{U}^\dag\boldsymbol{\Gamma \Gamma^\dag}\mathbf{U}$ such that the product,  \begin{equation}
  \sqrt{\epsilon}\mathbf{C}^\dag\frac{1}{\epsilon\boldsymbol{\mathcal{K}}} \mathbf{C}\sqrt{\epsilon}=\mathbf{I},
\end{equation}
gives an identity operator, and $\epsilon$ is a dimensionless parameter that controls the perturbation. Using this result, Eq.~\ref{Left equation Z factor} takes an alternative form,
\begin{equation}
 \begin{aligned}
 \frac{|b^j_\lambda|^2 }{|a^j_\lambda|^2}  &=\frac{1}{\boldsymbol{{Z}}_j (\omega_\lambda)}-1=\epsilon \bra{\boldsymbol{{\mathcal{R}}_j}}\mathbf{C}\frac{1}{\omega_{\lambda}-\mathbf{H}_{R|j}} \frac{1}{\omega_{\lambda}-\mathbf{H}_{R|j}} \mathbf{C}^\dag \ket{\boldsymbol{{\mathcal{R}}_j}}\\
  &= \frac{1}{\epsilon} \bra{\boldsymbol{{\mathcal{R}}_j}}\mathbf{C}\frac{1}{\omega_{\lambda}-\mathbf{H}_{R|j}} \mathbf{C}^\dag\epsilon  \frac{1}{\boldsymbol{\mathcal{K}}} \epsilon \mathbf{C} \frac{1}{\omega_{\lambda}-\mathbf{H}_{R|j}} \mathbf{C}^\dag \ket{\boldsymbol{{\mathcal{R}}_j}}=\frac{1}{\epsilon}\bra{\boldsymbol{{\mathcal{R}}_j}}  (\omega_{\lambda}-\mathbf{H}^{\boldsymbol{\mathcal{S}}}_j ) (\frac{1}{ \boldsymbol{\mathcal{K}}})(\omega_{\lambda}-\mathbf{H}^{\boldsymbol{\mathcal{S}}}_j )\ket{\boldsymbol{{\mathcal{R}}_j}}.\\
 \end{aligned}
 \label{overlap general}
\end{equation}
with a resolution of identity $ \mathbf{I}=\sum_{\mathcal{S}} \ket{\boldsymbol{{\mathcal{S}}_j}}\bra{\boldsymbol{{\mathcal{S}}_j}}$ in the projected Hilbert space, and we insert two identities into Eq.~\ref{overlap general}, 
\begin{equation}
\begin{aligned}
    \left. \frac{\partial \omega_{{\mathcal{R}}}(\omega) }{\partial \omega}\right|_{\omega_{\lambda }} &=  \frac{1}{\epsilon} \sum_{{\mathcal{S}}, {\mathcal{S}}'} \braket{\boldsymbol{{\mathcal{R}}_j}|\boldsymbol{{\mathcal{S}}_j}} \braket{\boldsymbol{{\mathcal{S}}_j}|\boldsymbol{{\mathcal{R}}_j}} (\omega_{\lambda}-\omega_{\mathcal{S}} ) (\frac{1}{\boldsymbol{\mathcal{K}}_{{\mathcal{S}} {\mathcal{S}}'}}) (\omega_{\lambda}-\omega_{{\mathcal{S}}'} ) \approx  \frac{1}{\epsilon}\mathbf{z_{j}}( \omega_\lambda) (\frac{1}{\boldsymbol{\mathcal{K}}_{{\mathcal{S}} {\mathcal{S}}}})(\omega_{\lambda}-\omega_{\mathcal{S}} )^2,\\
\end{aligned}
 \label{gradient equation}
\end{equation}
which recovers the Eq.~\ref{equality} in the main text. To the first order, the perturbation series has the form,
\begin{equation}
 \omega_\lambda \approx \omega_{\mathcal{S}} + \epsilon \boldsymbol{M_{{\mathcal{S}} {\mathcal{S}}}}(\omega_\lambda) +O(\epsilon^2), \hspace{0.2cm}  \ket{\boldsymbol{{\mathcal{R}}_j}} 
 \approx \ket{\boldsymbol{{\mathcal{S}}_j}} +  \epsilon \sum_{{\mathcal{S}}' \neq {\mathcal{S}}} \frac{ \boldsymbol{M_{{\mathcal{S}} {\mathcal{S}}'}} (\omega_\lambda)}{ \omega_{\mathcal{S}} -\omega_{{\mathcal{S}}'}}\ket{\boldsymbol{{\mathcal{S}}_j}}+O(\epsilon^2), \hspace{0.2cm} \boldsymbol{M_{{\mathcal{S}} {\mathcal{S}} }}(\omega_\lambda)=\braket{\boldsymbol{{\mathcal{S}}_j}|\boldsymbol{C} \frac{1}{\omega_{\mathcal{S}}-\mathbf{H}_{R|j}} \mathbf{C}^\dag|\boldsymbol{{\mathcal{S}}_j}}
\end{equation}
where $\epsilon$ is an expansion parameter. Following Eq.~\ref{gradient equation}, to the lowest order, the gradient has the form near the fixed point $\omega_{\mathcal{S}} +\epsilon\boldsymbol{M_{{\mathcal{S}} {\mathcal{S}}}}(\omega_\lambda) \approx \omega_{\lambda}$,
\begin{equation}
     \left. \frac{\partial \omega_{{\mathcal{R}}}(\omega) }{\partial \omega}\right|_{\omega_{\lambda }}   \approx \epsilon \sum^{\rm All}_{{\mathcal{S}}', {\mathcal{S}}''} \frac{ \boldsymbol{M_{{\mathcal{S}} {\mathcal{S}}'}}(\omega_\lambda) \boldsymbol{M _{{\mathcal{S}}'' {\mathcal{S}}}}(\omega_\lambda)}{{\boldsymbol{\mathcal{K}}_{{\mathcal{S}}' {\mathcal{S}}''}} } \approx \epsilon
 \frac{ (\boldsymbol{M_{{\mathcal{S}} {\mathcal{S}}}}(\omega_\lambda))^2  }{{\boldsymbol{\mathcal{K}}_{{\mathcal{S}} {\mathcal{S}}}} },
\end{equation}
where the last equality assumes  $ \boldsymbol{M _{{\mathcal{S}} {\mathcal{S}}}}^2/{\boldsymbol{\mathcal{K}}_{{\mathcal{S}} {\mathcal{S}}}} $ is dominate contribution to the gradient near the fixed point $\omega_\lambda$. The weight factor then takes the form,
\begin{equation}
     \boldsymbol{\mathcal{W}_j}( \omega_\lambda) \approx  \boldsymbol{{Z}_j}(\omega_\lambda)= 1-   \epsilon 
 \frac{ (\boldsymbol{M_{{\mathcal{S}} {\mathcal{S}}}}(\omega_\lambda))^2  }{{\boldsymbol{\mathcal{K}}_{{\mathcal{S}} {\mathcal{S}}}} }.
\end{equation}
By equating Eq.~\ref{grad curve} and Eq.~\ref{equality}, the similarity coefficient is related to the frequency variation near the fixed point $\omega_\lambda$, and as a sanity check, let's consider the approximation,
\begin{equation}
 \mathbf{z^{-1}_{j}}( \omega_\lambda) =
 \frac{\epsilon {\boldsymbol{\mathcal{K}}_{{\mathcal{S}} {\mathcal{S}}}}  }{\epsilon (\boldsymbol{M_{{\mathcal{S}} {\mathcal{S}}}}(\omega_\lambda))^2  }      \left. \frac{\partial \omega_{{\mathcal{R}}}(\omega) }{\partial \omega}\right|_{\omega_{\lambda }}  \approx  \frac{ {\boldsymbol{\mathcal{K}}_{{\mathcal{S}} {\mathcal{S}}}}  }{  (\boldsymbol{M_{{\mathcal{S}} {\mathcal{S}}}}(\omega_\lambda))^2  } 
 \frac{ (\boldsymbol{M_{{\mathcal{S}} {\mathcal{S}}}}(\omega_\lambda))^2  }{{\boldsymbol{\mathcal{K}}_{{\mathcal{S}} {\mathcal{S}}}} }=1+O(\epsilon^2).
    \label{z check}
\end{equation}

\textbf{\sffamily Hybridization function } the hybridization function has the form, 
\begin{equation}
\begin{aligned}
\lim_{\eta \to 0^+}\boldsymbol{\Delta}_{0}(\omega - i\eta)&=\boldsymbol{\mathcal{P}} \int_{-\infty}^{\infty} \mathrm{d}\omega' D_{0}(\omega')\frac{t_{0}^2(\epsilon)}{\omega-\omega'} + i\pi D_{0}(\omega) t_{0}^2(\omega),
\end{aligned}
\label{hybrid function}
\end{equation}
where $D_{0}(\omega')$ is density of states, and $t_{0}^2(\omega')$ is the coupling function to the continuum. Let's assume both the coupling and density of states are constants, i.e. $t_{0}(\omega')=t_{0}$ and $D_{0}(\omega)=D_{0}$. As a result, the hybridization function is an imaginary constant $|\Delta_0|= D_{0}  (t_{0})^2$ \cite{martin_reining_ceperley_2016}.

\textbf{\sffamily Numerical demonstration on two-level model } To demonstrate mutual renormalization effects between bath and system in the main text, we begin with a general Hamiltonian for a two-level system, 
\begin{equation}
\begin{aligned}
   H=&  
     \sum_{\sigma,\bar{\sigma}} \sum_{i}\omega_0 \lambda_i (c_{i,\sigma}^{\dag} c_{i,\sigma})
 +\sum_{ij} \omega_d (c_{i,\sigma}^{\dag} c_{j,\sigma}) +c.c. 
    +\sum_{ij} \left(V_{ij}n_{i,\sigma}n_{i,\Bar{\sigma}} 
   -J_{ij}c^\dag_{i,\bar{\sigma}}  c^\dag_{j,{\sigma}}  c_{j,\Bar{\sigma}}c_{i,\sigma} \right),
\end{aligned}
\label{Eq. sys Hamiltonian}
\end{equation}
which is characterized by an energy scale $\omega_0$, modulation parameter $\lambda_i$, the hopping strength $\mathrm{Re}\{\omega_d\}$, and the transition dipole moment strength $\mathrm{Im}\{\omega_d\}$. When $\mathrm{Im}\{\omega_d\} \equiv 0$, Eq.~\ref{Eq. sys Hamiltonian} processes a Hubbard-like model \cite{altland_simons_2010}, and when $\mathrm{Re}\{\omega_d\} \equiv 0$, it processes a Jaynes–Cummings-like model \cite{walls2008quantum}. The two-body terms, namely the density-density interaction parameter $V$ and the exchange interaction term $J$, characterize the local and non-local electron-electron correlation effects. We start with a dimer with site indices \textbf{0} and \textbf{x}, respectively. For simplicity, we focus on a system at half filling that shares opposite spins, e.g. $n=2, \uparrow\downarrow$, and the corresponding Hamiltonian in second quantization form is given by:  
\begin{equation}
\begin{aligned}
    H= & \sum_{\sigma}-\frac{3}{4}\omega_0  c_{0,\sigma}^{\dag} c_{0,\sigma} + \frac{\omega_d}{2} c_{x,\sigma}^{\dag} c_{0,\sigma}+\frac{ \omega^*_d}{2} c_{0,\sigma}^{\dag} c_{x,\sigma} +\frac{1}{4}\omega_0 c_{x,\sigma}^{\dag} c_{x,\sigma}\\
+&\sum_{\bar{\sigma}}V_{ij}n_{i,\sigma}n_{i,\Bar{\sigma}}
+J_{0x}(c^\dag_{0,\downarrow}  c^\dag_{x,\uparrow}  c_{x,\downarrow}c_{0,\uparrow} +c^\dag_{x,\downarrow}  c^\dag_{0,\uparrow}  c_{0,\downarrow}c_{x,\uparrow} ).
\end{aligned}
\end{equation}
  The $4\times 4$ Hamiltonian has a block form, 
\begin{equation}
        \mathbf{H} = \begin{bmatrix}
         \mathbf{H}^{\mathcal{S}}_{0\uparrow}& \mathbf{C}^\dag_{0x} \\
        \mathbf{C}_{0x} &\mathbf{H}_{x\uparrow} \\
    \end{bmatrix}, \hspace{0.4cm} \ket{\boldsymbol{\lambda}}=\begin{bmatrix}
        u_{00}\\ u_{0x}\\ u_{x0}\\ u_{xx}
    \end{bmatrix}=    u_{00} \ket{0\uparrow0\downarrow}+ u_{0x}\ket{0\uparrow x\downarrow}+ u_{x0} \ket{x\uparrow0\downarrow} + u_{xx}\ket{x\uparrow x\downarrow} ,
    \label{2 site H}
\end{equation}
where the eigenstates can be expressed as the linear combination of four basis states,
\begin{equation}
    \mathbf{H}^\mathcal{S}_{0\uparrow} = \begin{bmatrix}
        V_{00} -3\omega_0/2& \omega_d^*/2 \\
        \omega_d/2 & V_{0x}-\omega_0/2 \\
    \end{bmatrix}, \hspace{0.2cm} \mathbf{H}_{x\uparrow} = \begin{bmatrix}
        V_{0x} -\omega_0/2& \omega_d^*/2 \\
        \omega_d/2 & V_{xx}+\omega_0/2 \\
    \end{bmatrix}, \hspace{0.2cm} \mathbf{C}_{0x} = \begin{bmatrix}
       \omega_d/2& 0  \\
        J_{0x}& \omega_d/2\\
    \end{bmatrix}.
    \label{SI 2-site block H}
\end{equation}
 
The Pauli operators are given by the following, 
\begin{equation}
    \sigma^1=\begin{bmatrix}
        0,1\\
        1,0
    \end{bmatrix}, \hspace{0.2cm}    \sigma^2=\begin{bmatrix}
        0,-i\\
        i,0
    \end{bmatrix},\hspace{0.2cm}
      \sigma^3=\begin{bmatrix}
        1,0\\
        0,-1
    \end{bmatrix}. 
    \label{pauli}
\end{equation}
Under a unitary operation, a new orthogonal basis state can be obtained:
\begin{equation}
   \begin{aligned}
       & \ket{\text{bath, }(0,\phi,j)\uparrow}= \mathbf{U}_{j}\ket{\text{bath, }0\uparrow}=\psi_0\ket{\text{bath, }0\uparrow}+\psi_x\ket{\text{bath, }x\uparrow},\\
              & \ket{\text{bath, }(x,\phi,j)\uparrow}= \mathbf{U}_{j}\ket{\text{bath, }x\uparrow}=\psi_x\ket{\text{bath, }0\uparrow}-\psi_0\ket{\text{bath, }x\uparrow},
   \end{aligned} 
    \label{transformations}
\end{equation} 
where the coefficients are $\psi_0=\mathrm{cos}(\frac{\phi}{2})-i (\mathbf{n}_j)_z\mathrm{sin}(\frac{\phi}{2})$ and $
\psi_x=[(\mathbf{n}_j)_y-i (\mathbf{n}_j)_x ]\mathrm{sin}(\frac{\phi}{2})$. For each generated bath state $\ket{\text{bath, }(0,\phi,j)\uparrow}$, we obtain the corresponding degree of separability $\boldsymbol{{Z}_j}$ for all 4 eigenstates, as depicted in Fig.\ref{fig smooth curve} 1-3A.

\begin{figure*}[hbt!]
    \centering
    \includegraphics{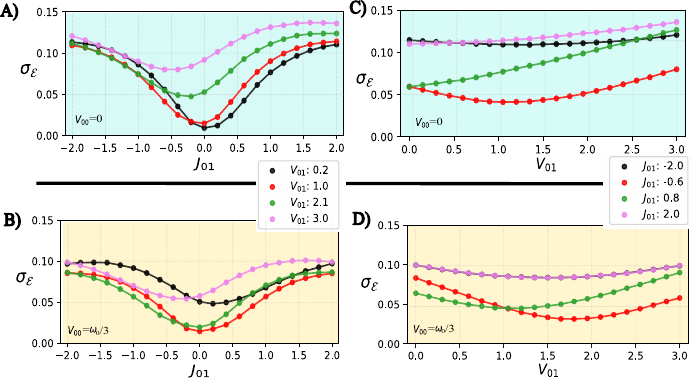}
    \caption{Standard deviation curves, discussed in Fig.~\ref{fig different V00} with identical labels.  }
    \label{fig entanglement_std}
\end{figure*}

\end{document}